\documentclass[twocolumn, journal]{IEEEtran}
\IEEEoverridecommandlockouts

\usepackage{color}
\usepackage{bm}
\usepackage{graphicx}
\usepackage{amsmath}
\usepackage{amssymb}
\usepackage{algorithm}
\usepackage{algorithmic}
\usepackage{multirow}
\usepackage{booktabs}
\usepackage{array}
\usepackage{lipsum}
\usepackage{enumerate}
\usepackage{stfloats}
\usepackage{subfigure}
\usepackage{cases}
\usepackage{diagbox}
\usepackage{cite}
\usepackage{enumitem}
\usepackage{algorithm,algorithmic}
\usepackage{balance}
\usepackage{amsthm}

\theoremstyle{plain}

\newtheorem{remark}{Remark}
\newtheorem{prop}{Proposition}

\newtheorem{theorem}{Theorem}
\newtheorem{lemma}{Lemma}

\allowdisplaybreaks[4]

\begin{document}

\title{Synergizing Covert Transmission and mmWave ISAC for Secure IoT Systems}

\author{
        Lingyun Xu, \IEEEmembership{Graduate Student Member~IEEE,}
        Bowen Wang, \IEEEmembership{Graduate Student Member~IEEE,} \\
        and Ziyang Cheng,~\IEEEmembership{Senior Member~IEEE}
       \vspace{-2em}

\thanks{The earlier version of this work was accepted and will be presented in part at the 2025 IEEE International Conference on Communication (ICC).}
\thanks{L. Xu and Z. Cheng are with the School of Information and Communication Engineering, University of Electronic Science and Technology of China, Chengdu 611731, China. (email: xusherly@std.uestc.edu.cn, zycheng@uestc.edu.cn).}
\thanks{B. Wang is with the King's Communications, Learning and Information Processing lab, King's College London, London, WC2R 2LS, UK. (email: bwwang@ieee.org)}
}

\maketitle

\begin{abstract}

This work focuses on the synergy of physical layer covert transmission and millimeter wave (mmWave) integrated sensing and communication (ISAC) to improve the performance, and enable secure internet of things (IoT) systems.
Specifically, we employ a physical layer covert transmission as a prism, which can achieve simultaneously transmitting confidential signals to a covert communication user equipment (UE) in the presence of a warden and regular communication UEs.
We design the transmit beamforming to guarantee information transmission security, communication quality-of-service (QoS) and sensing accuracy.
By considering two different beamforming architectures, i.e., fully digital beamforming (FDBF) and hybrid beamforming (HBF), an optimal design method and a low-cost beamforming scheme are proposed to address the corresponding problems, respectively.
Furthermore, building on the previously derived algorithm, two robust variants are proposed to address the more challenging case where the warden's CSI is imperfect.
Numerical simulations validate the effectiveness and superiority of the proposed FDBF/HBF algorithms compared with traditional algorithms in terms of information transmission security, communication QoS and target detection performance.

\end{abstract}


\section{Introduction}


\IEEEPARstart{T}{he} internet of things (IoT) in the future is expected to achieve interconnectivity between ubiquitous IoT devices \cite{cui2021integrating}.
This enables many emerging IoT applications, i.e., autonomous vehicles \cite{yurtsever2020survey}, environmental monitoring \cite{li2020review}, and smart home \cite{stojkoska2017review}, which need both wireless communication with high quality-of-service (QoS) and sensing with high accuracy.
To this end, integrated sensing and communication (ISAC) is proposed and envisioned as a key technology in the IoT, which integrates communication and sensing functionalities in a common wireless system by sharing the same frequency band and hardware resources\cite{liu2022integrated,lu2024integrated}.

Although ISAC has many advantages, the inherent broadcasting characteristic of wireless media gives rise to challenges for ISAC in information transmission security \cite{wei2022toward,qu2024privacy,shi2022intelligent}.
Previously, the investigations of information transmission security were mainly conducted on cryptography, where complicated encryption protocols, usernames, and passwords are created to guarantee the privacy of communication users \cite{diffie2022new}.
However, the security performance of cryptography is limited to complex secret key generation and management processes, compromised secret key identification, and eavesdroppers with strong computation capability \cite{wei2022toward}.

As a complementary approach to cryptography, physical layer security (PLS) is proposed to guarantee wireless transmission security by exploiting transmit degrees of freedom based on information theory \cite{wei2022toward,qu2024privacy,shi2022intelligent}.
Early works on PLS focused on the secrecy capability of the propagation channel \cite{mukherjee2014principles,wu2018survey}.
For example, authors in \cite{cumanan2013secrecy} investigated PLS for the multiple-input and multiple-output (MIMO) communication systems, where several secrecy rate optimization problems are solved in different scenarios.
Moreover, authors in \cite{su2020secure} studied the PLS in an ISAC system with a radar target acted as the eavesdropper, where artificial noise (AN) was adopted to congest the eavesdropper and the secrecy rate of the system was enhanced.
Another work in \cite{trigui2021secrecy} analyzed the PLS of reconfigurable intelligent surface (RIS) assisted communication systems, where the secrecy outage probability was derived in the presence of cooperative and separate eavesdroppers, respectively.
Although the works \cite{cumanan2013secrecy,su2020secure,trigui2021secrecy} can guarantee the PLS by reducing the amount of information intercepted by malicious eavesdroppers to some degree, the information transmission security would be vulnerable if the eavesdroppers have stronger decoding and computational capabilities.

Recently, unlike typical PLS techniques in \cite{cumanan2013secrecy,su2020secure,trigui2021secrecy}, a new research pioneered by Boulat A. Bash \textit{et al.} focuses on physical layer covert transmission \cite{bash2013limits}.
Specifically, the purpose of physical layer covert transmission is to prevent the wardens from detecting the existence of information transmission between communication users \cite{bash2013limits,makhdoom2022comprehensive,chen2023covert,jiang2024physical}.
By preventing the detection of information transmission, physical layer covert transmission can realize relatively better information transmission security than the typical PLS techniques, because the wardens would regard the covertly transmitted signals as the noises in the environment and thus would not try to tackle them. 
There have been some works on physical layer covert transmission for ISAC systems \cite{ma2022covert,liu2024exploiting,wang2024sensing}.
For instance, a covert beamforming design framework for ISAC systems was proposed in \cite{ma2022covert}, where the radar detection mutual information was maximized subject to the covertness constraint.
Another work in \cite{liu2024exploiting} investigated the ISAC system by simultaneously transmitting and reflecting reconfigurable intelligence surface (STAR-RIS), where the covert rate was optimized with imperfect channel state information.
Moreover, authors in \cite{wang2024sensing} investigated the physical layer covert transmission in a radar-communication cooperation system, where the radar was designed to jam and track the warden to improve the transmission covertness between the base station (BS) and a covert user.
Although the works \cite{ma2022covert,liu2024exploiting,wang2024sensing} can achieve satisfactory physical layer covert transmission in ISAC systems, they only consider a simple communication model involving one covert user and one warden, which is usually inapplicable in practical scenarios.

To serve more communication users and achieve communication with higher QoS, millimeter wave (mmWave) has attracted the attention of many researchers \cite{pi2011introduction}.
Moreover, MIMO also enables higher resolution and sensitivity for radar sensing \cite{li2007mimo}.
Based on these facts, mmWave and MIMO are promising to further enhance communication and sensing performances in ISAC systems.
Albeit with various merits, mmWave frequency suffers from harsh path losses \cite{heath2016overview}, so the mmWave MIMO systems exploit antenna arrays to conduct directional transmission by beamforming to compensate for the path losses.
However, the aforementioned works \cite{ma2022covert,liu2024exploiting,wang2024sensing} adopted the conventional fully digital beamforming (FDBF) structure, where every antenna requires one distinct radio frequency (RF) chain, leading to high power consumption and costs of RF chains and other hardware components, i.e., digital-to-analog converters (DAC).
To address these issues, a hybrid beamforming (HBF) structure was adopted in \cite{wang2021covert,ci2021hybrid,bai2023covert}, where a small number of RF chains are used to realize small-dimensional digital beamforming and a large number of phase shifters are used to realize large-dimensional analog beamforming \cite{alkhateeb2015limited,alkhateeb2014channel,ahmed2018survey}.
The physical layer covert transmission for full-duplex communications was investigated in \cite{wang2021covert},  where the HBF was designed by maximizing the covert rate.
Then, authors in \cite{ci2021hybrid} considered multiple covert users and proposed an HBF design for a multicast communication system, where the minimum covert rate was maximized subject to transmission covertness constraint.
Furthermore, authors in \cite{bai2023covert} analyzed the physical layer covert transmission for MIMO communication system with HBF structure, where the bounds on covert rates were derived under total variation distance and the Kullback-Leibler divergence covertness measures.
However, the works \cite{wang2021covert,ci2021hybrid,bai2023covert} only investigate physical layer covert transmission in communication systems.
To the best of our knowledge, hybrid beamforming for physical layer covert transmission in ISAC systems still remains an open problem, especially for a more general case with multiple regular users except for a covert user and a warden.

Motivated by the facts mentioned above, this paper investigates how to synergize covert transmission and mmWave ISAC to guarantee IoT security.
Specifically, the main contributions of this work are summarized as follows.

\begin{itemize}
    \item \textbf{Proposing Covert Transmission aided ISAC.} We propose a physical layer covert transmission aided secure mmWave ISAC for IoT systems, where a dual function BS (DFBS) aims to detect an IoT sensing target in the presence of clutters while simultaneously transmitting confidential data to a covert IoT communication user equipment (UE) in the presence of a warden and several regular communication IoT UEs.

    \item \textbf{Formulating Unified Design Problem.} We derive a unified optimization problem, where the covert rate is maximized while meeting the overt communication QoS, transmission covertness, and target detection requirements. 
    The proposed unified optimization problem accommodates both FDBF and HBF by adjusting the constraints and the form general of beamforming vectors, providing a flexible and systematic approach to analyzing different beamforming architectures.

    \item \textbf{Developing design Algorithms.} By enforcing the FDBF structure, we derive an optimal algorithm with theoretical optimality guarantees, which is based on the semidefinite relaxation (SDR). 
    In contrast, by enforcing the HBF structure, we develop a low-complexity algorithm based on alternating optimization.

    \item \textbf{Extending design with imperfect warden's CSI.} We further consider the scenario where the DFBS only has the imperfect warden's CSI due to non-cooperation between the DFBS and warden.
    To tackle this scenario, we extend the design methods into robust ones under FDBF and HBF architectures, respectively.

    \item \textbf{Providing insights and numerical validation.} We provide simulation results to evaluate the performance of the proposed FDBF and HBF schemes.
    The results demonstrate that, benefiting from the tailored design, the proposed FDBF and HBF algorithms outperform conventional approaches.
    It is also shown that while FDBF achieves superior performance, HBF offers greater hardware efficiency.
    Furthermore, the efficiency of the proposed covert transmission aided scheme is verified.
    
\end{itemize}

\begin{figure}[!t]
	\centering  \includegraphics[width=0.7\linewidth]{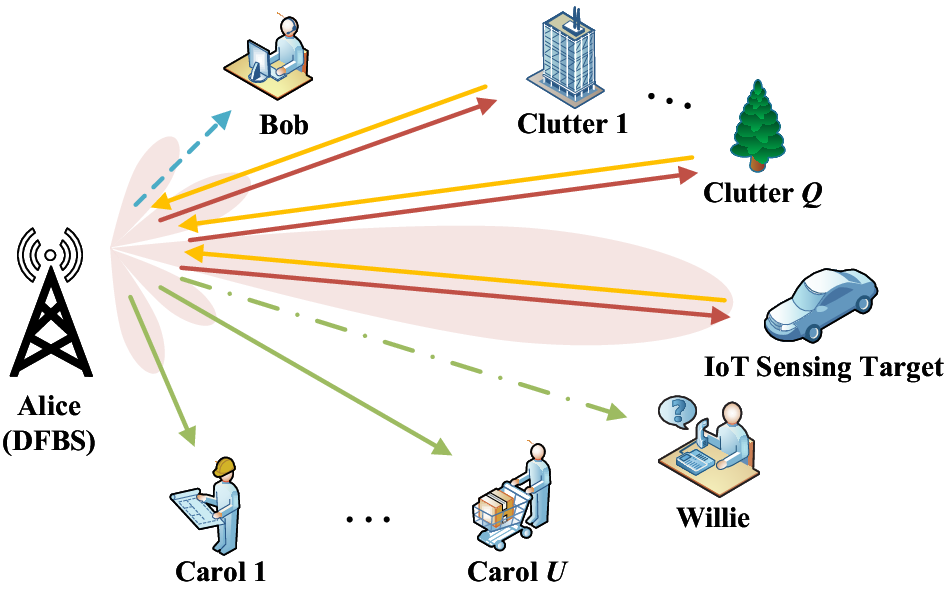}
	\vspace{-0.8em}
	\caption{An illustration of the proposed physical layer covert transmission aided mmWave ISAC for a secure IoT system.}
	\label{fig:scene_graph}
	\vspace{-1.5em}
\end{figure}

\textit{Organization:}
Section II introduces the system model and formulates the unified design problem.
Section III proposes the FDBF design algorithm and Section IV proposes the HBF design algorithm.
Section V demonstrates the numerical simulations and Section VII concludes this work.

\textit{Notations:} This paper uses $a$ for scalars, $\mathbf{a}$ for vectors and $\mathbf{A}$ for matrices. For a matrix $\mathbf{A}$, the element in the $i$-th row and the $j$-th column is denoted by ${\bf{A}}(i,j)$. $\mathbb{C}^{n}$ and $\mathbb{C}^{m\times n}$ denote an $n$ dimensional complex-valued vector and an $m$ by $n$ dimensional complex-valued space, respectively. $(\cdot)^T$ and $(\cdot)^H$ denote transpose and conjugate transpose operators, respectively. $\left|  \cdot  \right|$ represents a determinant or absolute value relying on the context. ${\left\|  \cdot  \right\|_F}$ and ${\rm Tr}({\cdot})$ denote the Frobenius norm and trace, respectively. ${\mathbb E}\left\{   \cdot \right\}$ denotes expectation. $\Re \left\{  \cdot  \right\}$  denotes the real part of a complex-valued number. $\mathcal{C}\mathcal{N}\left( {0,{\mathbf{R}}} \right)$ denotes the zero-mean complex Gaussian distribution with covariance matrix ${\mathbf{R}}$.
$\Pr \left( \mathcal A \right)$ denotes the probability of event $\mathcal A$ occurring.
${\bf{I}}_a$ denotes the $a \times a$ identity matrix, ${\bf{0}}_a$ and ${\bf{0}}_{a \times b}$ respectively denote the $a \times a$ and $a \times b$ matrices with all elements equal to 0.
${\rm vec}\left(\cdot\right)$ denotes vectorization.
${\rm{blkdiag}}\left(\cdot \right)$ denotes forming the block diagonal matrix from the input.

\section{System Model and Problem Formulation}\label{Sec-II}

\subsection{System Senario}

As shown in Fig. \ref{fig:scene_graph}, we consider a secure IoT system, where a DFBS (Alice) emits a probing waveform to detect an IoT sensing target in the presence of clutters while simultaneously serving overt IoT communication UEs and sending confidential communication data to a covert IoT UE (Bob).
Among these overt UEs, there are $U$ regular UEs (Carols) and one warden (Willie) who is hostile to Bob.
Specifically, Willie aims to determine whether the communication between Alice and Bob exists or not based on its received signals.
The sensing receiver co-located with the transmitter at Alice collects echo signals to detect the IoT sensing target.
Besides, to distinguish different IoT UEs, we define ${\cal R} = \left\{1,\dots,U\right\}$ as the collection of regular UEs corresponding to Carols, ${\cal P} = \left\{{\cal R},U+1\right\}$ as the collection of overt UEs corresponding to Carols and Willie, ${\cal A} = \left\{{\cal P},U+2 \right\}$ as the total collection of UEs corresponding to Carols, Willie and Bob.

\subsection{Transmit Signal}

Suppose that Alice generates three types of communication data streams, i.e., ${{\bf{s}}_{\rm{C}}} = {\left[ {{s_{\rm{1}}}, \ldots ,{s_U}} \right]^T} \in {{\mathbb C}^U}$, ${{s_{\rm{W}}}}$ and ${{s_{\rm{B}}}}$.
Specifically, ${{\bf{s}}_{\rm{C}}}$ denotes the data symbols for Carols, ${{s_{\rm{W}}}}$ denotes the data symbol for Willie, and ${{s_{\rm{B}}}}$ denotes the data symbol for Bob, assuming ${\mathbb E}\left\{ {{{\bf{s}}}{\bf{s}}^H} \right\} = {{\bf{I}}_{U+2}}$ with ${\bf s} = \left[{\bf s}_{\rm C}^T,{s_{\rm{W}}},{s_{\rm{B}}}\right]^T$.
These data streams are processed by the beamformer and the resultant ISAC signals are emitted by the transmit antennas from Alice.

Since Alice needs to transmit communication signals to the overt UEs, Carols and Willie, and intermittently transmit confidential communication signals to Bob, there is a binary detection hypothesis for the transmit signal from Alice.
Specifically, the general form of the transmit signal is given by
\begin{equation}\label{eq:1}
	{\bf{x}} = \left\{ \begin{array}{l}
	{\cal H}{_0}:{{\bf{V}}_{\rm{C}}}{{\bf{s}}_{\rm{C}}} + {{\bf{v}}_{\rm{W}}}{s_{\rm{W}}},\\
	{\cal H}{_1}:{{\bf{v}}_{\rm{B}}}{s_{\rm{B}}} + {{\bf{V}}_{\rm{C}}}{{\bf{s}}_{\rm{C}}} + {{\bf{v}}_{\rm{W}}}{s_{\rm{W}}},
	\end{array} \right.
\end{equation} 
where ${\cal H}{_0}$ denotes the null hypothesis that Alice only transmits overt signals to Carols and Willie, and ${\cal H}{_1}$ denotes the alternative hypothesis that Alice transmits overt signals to Carols and Willie while simultaneously transmitting covert signal to Bob.
Besides, ${{\bf{V}}_{\rm{C}}} = [ {{{\bf{v}}_{{\rm{C}},1}}, \ldots ,{{\bf{v}}_{{\rm{C}},U}}} ] \in {\mathbb C}^{M_t \times U}$, ${{\bf{v}}_{\rm{W}}} \in {\mathbb C}^{M_t}$, and ${{\bf{v}}_{\rm{B}}} \in {\mathbb C}^{M_t}$ denotes the general form of the beamforming vectors for Carols, Willie and Bob, respectively.

\subsection{Communication Model and Performance Metric}

Under the two hypotheses mentioned above, the transmit signal \eqref{eq:1} propagates through the mmWave channel, and the received signals at the $u$-th Carol, Willie and Bob can be respectively expressed as
\begin{align}
&\!\!\!\!{y_{{\rm{C}},u}} 
=\left\{ \begin{array}{l}
{\cal H}{_0}:{\bf{h}}_{{\rm{C}},u}^H\left( {{{\bf{V}}_{\rm{C}}}{{\bf{s}}_{\rm{C}}} + {{\bf{v}}_{\rm{W}}}{s_{\rm{W}}}} \right) + {n_{{\rm{C}},u}},\\
{\cal H}{_1}:{\bf{h}}_{{\rm{C}},u}^H\left( {{{\bf{v}}_{\rm{B}}}{s_{\rm{B}}} + {{\bf{V}}_{\rm{C}}}{{\bf{s}}_{\rm{C}}} + {{\bf{v}}_{\rm{W}}}{s_{\rm{W}}}} \right) + {n_{{\rm{C}},u}},
\end{array} \right.\label{eq:2}\\
&{y_{\rm{W}}} 
 = \left\{ \begin{array}{l}
{\cal H}{_0}:{\bf{h}}_{\rm{W}}^H\left( {{{\bf{V}}_{\rm{C}}}{{\bf{s}}_{\rm{C}}} + {{\bf{v}}_{\rm{W}}}{s_{\rm{W}}}} \right) + {n_{\rm{W}}},\\
{\cal H}{_1}:{\bf{h}}_{\rm{W}}^H\left( {{{\bf{v}}_{\rm{B}}}{s_{\rm{B}}} + {{\bf{V}}_{\rm{C}}}{{\bf{s}}_{\rm{C}}} + {{\bf{v}}_{\rm{W}}}{s_{\rm{W}}}} \right) + {n_{\rm{W}}},
\end{array} \right.\label{eq:3}\\
&{y_{\rm{B}}} 
= \left\{ \begin{array}{l}
{\cal H}{_0}:{\bf{h}}_{\rm{B}}^H\left( {{{\bf{V}}_{\rm{C}}}{{\bf{s}}_{\rm{C}}} + {{\bf{v}}_{\rm{W}}}{s_{\rm{W}}}} \right) + {n_{\rm{B}}},\\
{\cal H}{_1}:{\bf{h}}_{\rm{B}}^H\left( {{{\bf{v}}_{\rm{B}}}{s_{\rm{B}}} + {{\bf{V}}_{\rm{C}}}{{\bf{s}}_{\rm{C}}} + {{\bf{v}}_{\rm{W}}}{s_{\rm{W}}}} \right) + {n_{\rm{B}}},
\end{array} \right.\label{eq:4}
\end{align}
where ${{\bf{h}}_{{\rm{C,}}u}}, {\bf{h}}_{\rm{W}}, {{\bf{h}}_{\rm{B}}} \in {{\mathbb C}^{M_t}}$ denotes the channel vectors for the communication links Alice-to-Carol $u$, Alice-to-Willie, Alice-to-Bob, respectively.
Besides, ${n_{{\rm{C}},u}} \sim {\cal C}{\cal N}\left( {0,{\sigma _{{\rm{C}},u}^2}} \right),{n_{\rm{W}}} \sim {\cal C}{\cal N}\left( {0,{\sigma _{\rm{W}}^2}} \right),{n_{\rm{B}}^2} \sim {\cal C}{\cal N}\left( {0,{\sigma _{\rm{B}}^2}} \right)$ denotes the corresponding additive white Gaussian noise (AWGN) of these communication links.
To analyze communication performance more specifically, we characterize overt communication QoS and transmission covertness in the following.

\subsubsection{Communication QoS}
Since this work focuses on investigating physical layer covert transmission in the considered IoT system, we characterize the communication performance in the case where the communication links Alice-to-Carols, Alice-to-Willie, and Alice-to-Bob simultaneously exist.
In this case, based on the received signals \eqref{eq:2}-\eqref{eq:4}, the achievable communication data rates at the $u$-th Carol, Willie and Bob can be respectively calculated by
\begin{align}
&{{\rm Rate}_{{\rm{C}},u}} = \log_2 \left( {1 + {\rm{SIN}}{{\rm{R}}_{{\rm{C}},u}}} \right),\label{eq:5}\\
&{{\rm Rate}_{\rm{W}}} = \log_2 \left( {1 + {\rm{SIN}}{{\rm{R}}_{\rm{W}}}} \right), \label{eq:6}\\
&{{\rm Rate}_{\rm{B}}} = \log_2 \left( {1 + {\rm{SIN}}{{\rm{R}}_{\rm{B}}}} \right), \label{eq:7}
\end{align}
where the signal-to-interference-plus-noise ratios (SINRs) at the $u$-th Carol, Willie and Bob are respectively given by
\begin{align}
&{\rm{SIN}}{{\rm{R}}_{{\rm{C}},u}} = \frac{{{{\left| {{\bf{h}}_{{\rm{C}},u}^H{{\bf{v}}_{{\rm{C,}}u}}} \right|}^2}}}{{\sum\limits_{\scriptstyle i \in {\cal R}\hfill\atop
\scriptstyle i \ne u\hfill} \!\!{{{\left| {{\bf{h}}_{{\rm{C}},u}^H{{\bf{v}}_{{\rm{C}},i}}} \right|}^2}} \!\! + \! {{\left| {{\bf{h}}_{{\rm{C}},u}^H{{\bf{v}}_{\rm{W}}}} \right|}^2} \!\! + \! {{\left| {{\bf{h}}_{{\rm{C}},u}^H{{\bf{v}}_{\rm{B}}}} \right|}^2} \!\!+\! \sigma _{{\rm{C}},u}^2}}, \nonumber\\
&{\rm{SIN}}{{\rm{R}}_{\rm{W}}} =  \frac{{{{\left| {{\bf{h}}_{\rm{W}}^H{{\bf{v}}_{\rm{W}}}} \right|}^2}}}{{\sum\limits_{i \in {\cal R}} {{{\left| {{\bf{h}}_{\rm{W}}^H{{\bf{v}}_{{\rm{C}},i}}} \right|}^2}}  + {{\left| {{\bf{h}}_{\rm{W}}^H{{\bf{v}}_{\rm{B}}}} \right|}^2} + \sigma _{\rm{W}}^2}}, \nonumber\\
&{\rm{SIN}}{{\rm{R}}_{\rm{B}}} =  \frac{{{{\left| {{\bf{h}}_{\rm{B}}^H{{\bf{v}}_{\rm{B}}}} \right|}^2}}}{{\sum\limits_{i \in {\cal R}} {{{\left| {{\bf{h}}_{\rm{B}}^H{{\bf{v}}_{{\rm{C}},i}}} \right|}^2}}  + {{\left| {{\bf{h}}_{\rm{B}}^H{{\bf{v}}_{\rm{W}}}} \right|}^2} + \sigma _{\rm{B}}^2}}.\nonumber
\end{align}
To distinguish the achievable communication data rates at the overt communication UEs and the covert communication UE, we call ${\rm Rate}_{{\rm C},u}, \forall u$ and ${\rm Rate}_{\rm W}$ as the overt communication rates and ${\rm Rate}_{\rm B}$ as the covert communication rate.

\subsubsection{Transmission Covertness}

Since Willie wants to determine whether the communication link Alice-to-Bob exists based on its received signal \eqref{eq:3}, it is necessary to analyze the probability distribution of $y_{\rm W}$.
To this end, from the perspective of Willie, we define the likelihood function ${\mathbb P}_0$ for the case where ${\mathcal H}_0$ is true and ${\mathbb P}_1$ for the case where ${\mathcal H}_1$ is true.
Specifically, the corresponding likelihood functions can be expressed as
\begin{equation}\label{eq:8}
	\left\{ \begin{array}{l}
	{{\mathbb P}_0}:p\left( {{y_{\rm{W}}}|{{\cal H}_0}} \right) = \frac{1}{{\pi {\kappa _0}}}\exp \left( { - \frac{{{{\left| {{y_{\rm{W}}}} \right|}^2}}}{{{\kappa _0}}}} \right),\\
	{{\mathbb P}_1}:p\left( {{y_{\rm{W}}}|{{\cal H}_1}} \right) = \frac{1}{{\pi {\kappa _1}}}\exp \left( { - \frac{{{{\left| {{y_{\rm{W}}}} \right|}^2}}}{{{\kappa _1}}}} \right),
	\end{array} \right.
\end{equation}
where ${\kappa _0} = \sum\nolimits_{i \in {\cal R}} {{{| {{\bf{h}}_{\rm{W}}^H{{\bf{v}}_{{\rm{C}},i}}} |}^2}}  + {| {{\bf{h}}_{\rm{W}}^H{{\bf{v}}_{\rm{W}}}} |^2} + \sigma _{\rm{W}}^2$ and ${\kappa _1} = {| {{\bf{h}}_{\rm{W}}^H{{\bf{v}}_{\rm{B}}}} |^2} + \sum\nolimits_{i \in {\cal R}} {{{| {{\bf{h}}_{\rm{W}}^H{{\bf{v}}_{{\rm{C}},i}}} |}^2}}  + {| {{\bf{h}}_{\rm{W}}^H{{\bf{v}}_{\rm{W}}}} |^2} + \sigma _{\rm{W}}^2$.

When distinguishing two hypothesises ${\mathcal H}_0$ and ${\mathcal H}_1$, due to the randomness of the received signal, it is inevitable for Willie to have two types of detection errors: 1) \textit{false alarm}: Willie makes the decision ${\mathcal D}_1$ while the hypothesis ${\mathcal H}_0$ holds; 2) \textit{missed detection}: Willie makes the decision ${\mathcal D}_0$ while the hypothesis ${\mathcal H}_1$ holds, where ${\mathcal D}_0$ and ${\mathcal D}_1$ denote the binary decisions corresponding to the hypothesises ${\mathcal H}_0$ and ${\mathcal H}_1$, respectively.
Based on this, the false alarm probability is defined as $\Pr \left( {{{\cal D}_1}|{{\cal H}_0}} \right)$ and the missed detection probability is defined as $\Pr\left( {{{\cal D}_0}|{{\cal H}_1}} \right)$.
Suppose Willie adopts the classical hypothesis test with equal prior probabilities of the hypothesises ${\mathcal H}_0$ and ${\mathcal H}_1$, the total detection error probability at Willie, $P_{\rm E}$, can be calculated by
\begin{equation}\label{eq:9}
P_{\rm E} = \Pr \left( {{{\cal D}_1}|{{\cal H}_0}} \right) + \Pr\left( {{{\cal D}_0}|{{\cal H}_1}} \right).
\end{equation}

To be consistent with the conventional statistical hypothesis test in detection, we assume that an optimal statistical hypothesis test, Neyman-Pearson (NP) test \cite{lehmann1986testing} of minimizing the total detection error probability $P_{\rm E}$, is constructed by Willie.
Specifically, the corresponding likelihood ratio test is given by
\begin{equation}\label{eq:10}
\frac{{p\left( {{y_{\rm{W}}}|{{\cal H}_1}} \right)}}{{p\left( {{y_{\rm{W}}}|{{\cal H}_0}} \right)}} \underset{\mathcal{D}_0}{\overset{\mathcal{D}_1}{\gtrless} } 1.
\end{equation}
Then, we give the following proposition to calculate the false alarm probability $\Pr \left( {{{\cal D}_1}|{{\cal H}_0}} \right)$ and the missed detection probability $\Pr \left( {{{\cal D}_0}|{{\cal H}_1}} \right)$.
\begin{prop}\label{prop:1}
	Under an optimal detection rule in NP test, the false alarm probability $\Pr \left( {{{\cal D}_1}|{{\cal H}_0}} \right)$ and the missed detection probability $\Pr \left( {{{\cal D}_0}|{{\cal H}_1}} \right)$ are respectively given by
	\begin{align}
	    \Pr \left( {{{\cal D}_1}|{{\cal H}_0}} \right) =& {\left( {\frac{{{\kappa _1}}}{{{\kappa _0}}}} \right)^{ - \frac{{{\kappa _1}}}{{{\kappa _1} - {\kappa _0}}}}},\label{eq:11}\\
		\Pr \left( {{{\cal D}_0}|{{\cal H}_1}} \right) =& 1 - {\left( {\frac{{{\kappa _1}}}{{{\kappa _0}}}} \right)^{ - \frac{{{\kappa _0}}}{{{\kappa _1} - {\kappa _0}}}}}\label{eq:12}.
	\end{align}
\end{prop}
\begin{IEEEproof}
	Please refer to Appendix \ref{app:A}.
\end{IEEEproof}
Based on \textbf{Proposition \ref{prop:1}}, we substitute \eqref{eq:11} and \eqref{eq:12} into \eqref{eq:9} and obtain the total detection error probability at Willie, which can be expressed as
\begin{equation}\label{eq:13}
	P_{\rm E} = 1 + {\left( {\frac{{{\kappa _1}}}{{{\kappa _0}}}} \right)^{ - \frac{{{\kappa _1}}}{{{\kappa _1} - {\kappa _0}}}}}  - {\left( {\frac{{{\kappa _1}}}{{{\kappa _0}}}} \right)^{ - \frac{{{\kappa _0}}}{{{\kappa _1} - {\kappa _0}}}}}.
\end{equation}
Note that $P_{\rm E}$ should be as high as possible to guarantee the transmission covertness of Alice-to-Bob communication.

\subsection{Sensing Model and Performance Metric}

Assume that there is an IoT sensing target in the presence of $Q$ stationary clutter sources, i.e., trees and buildings, located at $\theta_0$ and $\theta_q, q \in {\mathcal Q} = \left\{ 1, \dots, Q\right\}$, respectively.
Given the transmitted signal \eqref{eq:1}, the echo signals collected by the sensing receiver co-located with the transmitter at Alice is given by
\begin{equation}\label{eq:14}
    {\bf{r}} = {\varsigma _0}{{\bf{a}}_r}\left( {{\theta _0}} \right){\bf{a}}_t^H\left( {{\theta _0}} \right){\bf{x}} + \sum\limits_{q \in {\cal Q}} {{\varsigma _q}{{\bf{a}}_r}\left( {{\theta _q}} \right){\bf{a}}_t^H\left( {{\theta _q}} \right){\bf{x}}}  + {\bf{z}},
\end{equation}
where ${\bf{z}} \sim {\cal C}{\cal N}\left( {{\bf{0}},\sigma _{\rm{r}}^2{{\bf{I}}_{M_r}}} \right)$ denotes the AWGN, $\varsigma_0$ and $\varsigma_q,\forall q$ denote the complex amplitude of the IoT sensing target and clutters, respectively.
${{\bf{a}}_t}(\theta )$ and ${{\bf{a}}_r}(\theta )$ denote the transmit and receive steering vectors, respectively, which are given by
\begin{equation}
\begin{aligned}
&{{\bf{a}}_t}(\theta ) = \frac{1}{{\sqrt {{M_t}} }}{\left[ {1,\;{e^{ \jmath \pi \sin \theta }}, \ldots ,\;{e^{ \jmath \pi \left( {{M_t} - 1} \right)\sin \theta }}} \right]^T} \in {{\mathbb C}^{{M_t}}}, \\
&{{\bf{a}}_r}(\theta ) = \frac{1}{{\sqrt {{M_r}} }}{\left[ {1,\;{e^{ \jmath \pi \sin \theta }}, \ldots ,\;{e^{ \jmath \pi \left( {{M_r} - 1} \right)\sin \theta }}} \right]^T} \in {{\mathbb C}^{{M_r}}} . \nonumber
\end{aligned}
\end{equation}

Then, the received signal \eqref{eq:14} is processed by a sensing receive filter ${\bf w} \in {\mathbb C}^{M_r}$, leading to the following binary hypothesis test for target detection
\begin{equation}\label{eq:15}
\begin{aligned}
{{\tilde r}} &= {{\bf{w}}^H}{\bf{r}} \\
& = \left\{ \begin{array}{l}
{\widetilde {\cal H}_0}:\sum\limits_{q \in {\cal Q}} {{\varsigma _q}{{\bf{w}}^H}{{\bf{a}}_r}\left( {{\theta _q}} \right){\bf{a}}_t^H\left( {{\theta _q}} \right){\bf{x}}}  + {{\bf{w}}^H}{\bf{z}},\\
{\widetilde {\cal H}_1}:{\varsigma _0}{{\bf{w}}^H}{{\bf{a}}_r}\left( {{\theta _0}} \right){\bf{a}}_t^H\left( {{\theta _0}} \right){\bf{x}}\\
\qquad  + \sum\limits_{q \in {\cal Q}} {{\varsigma _q}{{\bf{w}}^H}{{\bf{a}}_r}\left( {{\theta _q}} \right){\bf{a}}_t^H\left( {{\theta _q}} \right){\bf{x}}}  + {{\bf{w}}^H}{\bf{z}},
\end{array} \right.
\end{aligned}
\end{equation}
where ${\widetilde {\cal H}_0}$ denotes the null hypothesis that there is no IoT sensing target, and  ${\widetilde {\cal H}_1}$ denotes the alternative hypothesis that there exists an IoT sensing target.

Based on the binary hypothesis for target detection \eqref{eq:15}, we consider the generalized likelihood ratio test (GLRT) detector \cite{richards2005fundamentals} at the sensing receiver.
Given a false alarm probability $P_{fa}$, we can derive the target detection probability $P_{d}$ by
    \begin{equation}\label{eq:16}
        P_{d} = {Q_M}\left( {\sqrt {2{\rm{SIN}}{{\rm{R}}_{\rm{r}}}} ,\sqrt { - 2\ln {P_{fa}}} } \right),
    \end{equation}
    where ${Q_M}\left( \cdot ,\cdot \right)$ is the Marcum Q function with order 1, and the output sensing SINR, $\rm SINR_r$, is given in the equation \eqref{eq:17},
       \begin{figure*}[!ht]
    \begin{equation}\label{eq:17}
        {\rm{SIN}}{{\rm{R}}_{\rm{r}}} = \frac{{\sum\limits_{i \in {\cal R}} {{{\left| {{\varsigma _0}{{\bf{w}}^H}{\bf{A}}\left( {{\theta _0}} \right){{\bf{v}}_{{\rm{C}},i}}} \right|}^2}}  + {{\left| {{\varsigma _0}{{\bf{w}}^H}{\bf{A}}\left( {{\theta _0}} \right){{\bf{v}}_{\rm{W}}}} \right|}^2} + {{\left| {{\varsigma _0}{{\bf{w}}^H}{\bf{A}}\left( {{\theta _0}} \right){{\bf{v}}_{\rm{B}}}} \right|}^2}}}{{\sum\limits_{q \in {\cal Q}} {\left( {\sum\limits_{i \in {\cal R}} {{{\left| {{\varsigma _q}{{\bf{w}}^H}{\bf{A}}\left( {{\theta _q}} \right){{\bf{v}}_{{\rm{C}},i}}} \right|}^2}}  + {{\left| {{\varsigma _q}{{\bf{w}}^H}{\bf{A}}\left( {{\theta _q}} \right){{\bf{v}}_{\rm{W}}}} \right|}^2} + {{\left| {{\varsigma _q}{{\bf{w}}^H}{\bf{A}}\left( {{\theta _q}} \right){{\bf{v}}_{\rm{B}}}} \right|}^2}} \right)}  + \sigma _{\rm{r}}^2\left\| {\bf{w}} \right\|_F^2}},
    \end{equation}
    \hrule
    \vspace{-1em}
    \end{figure*}
where we define ${\bf A}\left(\theta\right) = {{\bf{a}}_r}\left( {{\theta}} \right){\bf{a}}_t^H\left( {{\theta}} \right)$.

Accordingly, we make the following remark.
\begin{remark}\label{remark:1}(Design Metric Conversion)
    It can be noticed that with a given false alarm probability $P_{fa}$, the target detection probability \eqref{eq:16} is a monotonically increasing function about the output sensing SINR, $\rm SINR_r$.
    Therefore, to guarantee a satisfactory target detection performance of the proposed IoT system, we can convert the problem of optimizing the target detection probability into the problem of optimizing the output sensing SINR.
\end{remark}

\subsection{Problem Formulation}

To realize covert transmission and mmWave ISAC in the considered IoT system, the transmit beamformer, $\{{\bf V}_{\rm C},{\bf v}_{\rm W},{\bf v}_{\rm B}\}$, and sensing receive filter, ${\bf w}$, should be jointly designed by maximizing the covert rate subject to overt communication QoS, transmission covertness, target detection and transmit power constraints.
Mathematically, the unified design problem can be formulated as
\begin{subequations}
    \begin{align}
        &\mathop {\max }\limits_{{{\bf{V}}_{{\rm{C}}}},{{\bf{v}}_{{\rm{W}}}},{{\bf{v}}_{{\rm{B}}}},{\bf{w}}} {\rm{Rate}}_{\rm{B}}\left({\bf V}_{\rm C},{\bf v}_{\rm W},{\bf v}_{\rm B}\right),\label{eq:P0-a}\\
        &\qquad\;{\rm{s.t.}}\quad\; \left\| {{{\bf{V}}_{{\rm{C}}}}} \right\|_F^2 + \left\| {{{\bf{v}}_{{\rm{W}}}}} \right\|_F^2 + \left\| {{{\bf{v}}_{{\rm{B}}}}} \right\|_F^2 \le P,\label{eq:P0-b}\\
        &\qquad\qquad\quad{\rm{Rate}}_{{\rm{C}},u}\left({\bf V}_{\rm C},{\bf v}_{\rm W},{\bf v}_{\rm B}\right) \ge {\xi _u},\;u \in {\cal R},\label{eq:P0-c}\\
        &\qquad\qquad\quad{\rm{Rate}}_{\rm{W}}\left({\bf V}_{\rm C},{\bf v}_{\rm W},{\bf v}_{\rm B}\right) \ge {\xi _{\rm{W}}},\label{eq:P0-d}\\
        &\qquad\qquad\quad1-{P_{\rm{E}}}\left({\bf V}_{\rm C},{\bf v}_{\rm W},{\bf v}_{\rm B}\right) \le \epsilon ,\label{eq:P0-e}\\
        &\qquad\qquad\quad{\rm{SINR}}_{\rm{r}}\left({\bf V}_{\rm C},{\bf v}_{\rm W},{\bf v}_{\rm B},{\bf w}\right) \ge \gamma,\label{eq:P0-f}\\
        &\qquad\qquad\quad {\bf V}_{\rm C},{\bf v}_{\rm W},{\bf v}_{\rm B} \in \mathcal{S},\label{eq:P0-g}
    \end{align}\label{eq:P0}%
\end{subequations}
where $P$ denotes the total transmit power, $\xi_u,\forall u$ and $\xi_{\rm W}$ denote the overt communication QoS requirements of Carols and Willie, respectively.
Besides, $\epsilon$ denotes the transmission covertness constraint, and $\gamma$ denotes the sensing SINR threshold.
It is worth mentioning that the specific form of the transmit beamforming vector $\{{\bf V}_{\rm C},{\bf v}_{\rm W},{\bf v}_{\rm B}\}$ and the additional constraint \eqref{eq:P0-g} are determined by the specific transmit structure, where $\mathcal{S}$ is the hardware constraint set.

\begin{figure}[!t]
   \centering
   \subfigure[]{
   \includegraphics[height=0.246\linewidth]{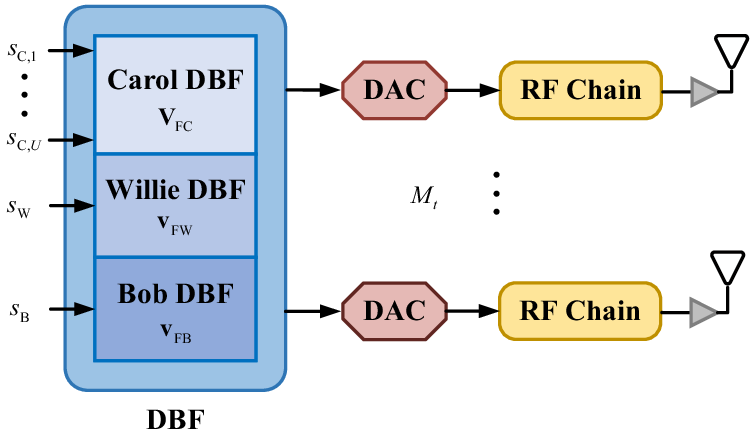} 
   \label{fig:FD}	}
   \hspace{-1.3em}
   \subfigure[]{
   \includegraphics[height=0.246\linewidth]{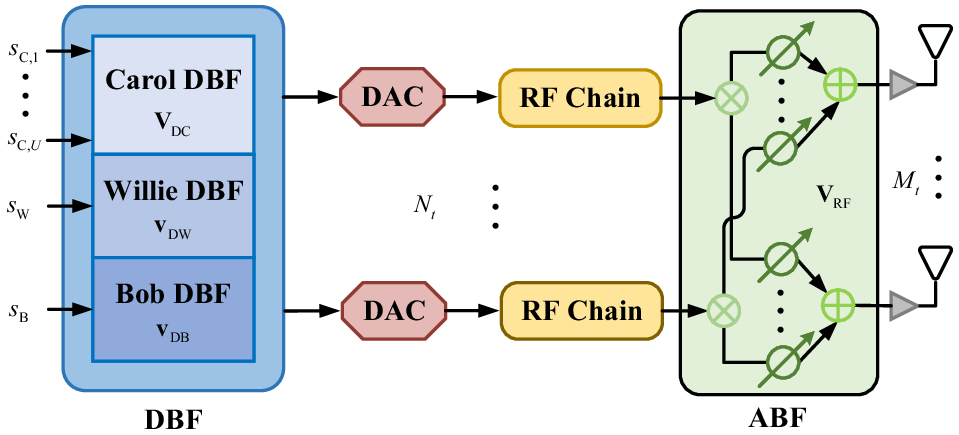} 
   \label{fig:HBF}	}
   \vspace{-1em}
   \caption{The proposed transmit structures at Alice: (a) FDBF structure; (b) HBF structure with fully-connected phase shift network.}
   \label{fig:BF_strut}
   \vspace{-1em}
\end{figure}

To solve the formulated design problem when adopting different transmit beamforming structures, we consider two transmit beamforming adopting FDBF and HBF as shown in Fig. \ref{fig:BF_strut}, and devise the corresponding algorithms to solve the corresponding specific problems in the next two sections.

\section{FDBF Design Algorithm}\label{sec:III}

In this section, we consider the case where the FDBF structure shown in Fig. \ref{fig:FD} is implemented in the transmitter at Alice and propose an optimal algorithm to jointly design the transmit FDBF and the sensing receive filter.

\subsection{Problem Concretization under FDBF Structure}

\subsubsection{Employment of FDBF Structure}
We propose a transmitter adopting the FDBF structure in Fig. \ref{fig:FD}.
Specifically, the data streams ${{{\bf{s}}_{\rm{C}}},{s_{\rm{W}}},{s_{\rm{B}}}}$ are processed by the Carol FDBF ${{{\bf{V}}_{\rm{FC}}} = \left[ {{{\bf{v}}_{{\rm{FC}},1}}, \ldots ,{{\bf{v}}_{{\rm{FC}},U}}} \right] \in {{\mathbb C}^{{M_t} \times U}}}$, Willie FDBF ${{\bf{v}}_{\rm{FW}}} \in {{\mathbb C}^{{M_t}}}$, and Bob FDBF ${{\bf{v}}_{\rm{FB}}} \in {{\mathbb C}^{{M_t}}}$, respectively.
Then, the processed signals after DACs are up-converted by $M_t$ RF chains.
Finally, the signals are emitted by $M_t$ antennas.
In this case, we substitute ${\bf V}_{\rm C} = {{\bf{V}}_{\rm{FC}}}$, ${\bf v}_{\rm W} = {{\bf{v}}_{\rm{FW}}}$ and ${\bf v}_{\rm B} = {{\bf{v}}_{\rm{FB}}}$ into the general form of the transmit signal \eqref{eq:1}.

\subsubsection{Problem Concretization}

Adopting the FDBF structure at the transmitter, the unified design problem \eqref{eq:P0} can be concretized to a joint design of the FDBF $\left\{{\bf V}_{\rm FC},{\bf v}_{\rm FW},{\bf v}_{\rm FB}\right\}$ and sensing receive filter $\bf w$, which is given by
\begin{subequations}
    \begin{align}
        &\mathop {\max }\limits_{{{\bf{V}}_{{\rm{FC}}}},{{\bf{v}}_{{\rm{FW}}}},{{\bf{v}}_{{\rm{FB}}}},{\bf{w}}} {\rm{Rat}}{{\rm{e}}_{\rm{B}}}\left({\bf V}_{\rm FC},{\bf v}_{\rm FW},{\bf v}_{\rm FB}\right),\label{eq:P1-a}\\
        &\qquad\;\;{\rm{s}}.{\rm{t}}.\quad \left\| {{{\bf{V}}_{{\rm{FC}}}}} \right\|_F^2 + \left\| {{{\bf{v}}_{{\rm{FW}}}}} \right\|_F^2 + \left\| {{{\bf{v}}_{{\rm{FB}}}}} \right\|_F^2 \le P,\label{eq:P1-b}\\
        &\qquad\qquad\quad{\rm{Rat}}{{\rm{e}}_{{\rm{C}},u}}\left({\bf V}_{\rm FC},{\bf v}_{\rm FW},{\bf v}_{\rm FB}\right) \ge {\xi _u},\;u \in {\cal R},\label{eq:P1-c}\\
        &\qquad\qquad\quad{\rm{Rat}}{{\rm{e}}_{\rm{W}}}\left({\bf V}_{\rm FC},{\bf v}_{\rm FW},{\bf v}_{\rm FB}\right) \ge {\xi _{\rm{W}}},\label{eq:P1-d}\\
        &\qquad\qquad\quad 1-{P_{\rm{E}}}\left({\bf V}_{\rm FC},{\bf v}_{\rm FW},{\bf v}_{\rm FB}\right) \le \epsilon ,\label{eq:P1-e}\\
        &\qquad\qquad\quad{\rm{SIN}}{{\rm{R}}_{\rm{r}}}\left({\bf V}_{\rm FC},{\bf v}_{\rm FW},{\bf v}_{\rm FB},{\bf w}\right) \ge \gamma,\label{eq:P1-f}
    \end{align}\label{eq:P1}%
\end{subequations}
which is non-convex due to the log-fractional expression in the objective function \eqref{eq:P1-a}.
To tackle this difficulty, we solve this problem in the next subsections.

\subsection{Receive Filter Optimization}\label{Sec-III-B-1}
To solve the problem \eqref{eq:P1}, we first notice that the receive filter $\bf w$ only exists in the constraint \eqref{eq:P1-f}.
Therefore, the receive filter optimization problem can be given by the following unconstrained problem
\begin{equation}\label{eq:19}
    \mathop {\max }\limits_{\bf{w}} \frac{{{{\bf{w}}^H}{\bf{\Xi }}_{\rm FD}{\bf{w}}}}{{{{\bf{w}}^H}{\bf{\Lambda }}_{\rm FD}{\bf{w}}}}
\end{equation}
where we define ${\bf{\Xi }}_{\rm FD} = {\left| {{\varsigma _0}} \right|^2}{\bf{A}}\left( {{\theta _0}} \right){{\bf{V}}_{{\rm{FD}}}}{\bf{V}}_{{\rm{FD}}}^H{{\bf{A}}^H}\left( {{\theta _0}} \right)$,  ${\bf{\Lambda }}_{\rm FD} = \sum\nolimits_{q \in {\cal Q}} {{{\left| {{\varsigma _q}} \right|}^2}{\bf{A}}\left( {{\theta _q}} \right){{\bf{V}}_{{\rm{FD}}}}{\bf{V}}_{{\rm{FD}}}^H{{\bf{A}}^H}\left( {{\theta _q}} \right)}  + \sigma _{\rm{r}}^2{{\bf{I}}_{{M_r}}}$,  ${{\bf{V}}_{{\rm{FD}}}} = \left[ {{{\bf{V}}_{{\rm{FC}}}},{{\bf{v}}_{{\rm{FW}}}},{{\bf{v}}_{{\rm{FB}}}}} \right] = \left[ {{{\bf{v}}_{{\rm{FD}},1}}, \ldots ,{{\bf{v}}_{{\rm{FD}},U + 2}}} \right] \in {{\mathbb C}^{{M_t} \times \left( {U + 2} \right)}}$ for simplicity of notation.

Note that the problem \eqref{eq:19} is a generalized Rayleigh quotient problem, whose optimal solution ${\bf w}^{\star}$ can be obtained by the generalized eigenvector corresponding to the maximum eigenvalue of ${( {{{\bf{\Lambda }}_{\rm FD}^{ - 1/2}}} )^H}{\bf{\Xi }}_{\rm FD}{{\bf{\Lambda }}_{\rm FD}^{ - 1/2}}$.

\subsection{Transmit FDBF Optimization}

\subsubsection{Problem Transformation}
It can be noticed that $P_{\rm E}$ contained in the constraint \eqref{eq:P1-e} involves complicated fractional and exponential functions, which is difficult to tackle.
To this end, we give the following lemma.
\begin{lemma}\label{lemma:1}
    Adopting the classical hypothesis test with the equal prior probabilities of ${\cal H}_0$ and ${\cal H}_1$ at Willie, the total detection error probability at Willie has the relation as follows.
    \begin{equation}\label{eq:20}
        P_{\rm E} = 1 - {{\cal V}_T}\left( {{{\mathbb P}_0},{{\mathbb P}_1}} \right),
    \end{equation}
    where ${{\cal V}_T}\left( {{{\mathbb P}_0},{{\mathbb P}_1}} \right) = \frac{1}{2}{\left\| {p\left( {{y_{\rm{W}}}|{{\cal H}_0}} \right) - p\left( {{y_{\rm{W}}}|{{\cal H}_1}} \right)} \right\|_F}$ is defined as the total variation distance between ${\mathbb P}_0$ and ${\mathbb P}_1$.
    Applying the Pinsker's inequality, we can further rewrite the lower bounded of $P_{\rm E}$ by
    \begin{equation}\label{eq:21}
        {P_{\rm{E}}} \ge 1 - \sqrt {\frac{1}{2}{\cal D}\left( {{{\mathbb P}_0}|{{\mathbb P}_1}} \right)}  \buildrel \Delta \over = {{\hat P}_{\rm{E}}},
    \end{equation}
    where ${\cal D}\left( {{{\mathbb P}_0}|{{\mathbb P}_1}} \right) = \ln ( {\frac{{{\kappa _1}}}{{{\kappa _0}}}} ) + \frac{{{\kappa _0}}}{{{\kappa _1}}} - 1$.
\end{lemma}
\begin{IEEEproof}
	Please refer to Appendix \ref{app:B}.
\end{IEEEproof}

Based on \textbf{Lemma \ref{lemma:1}}, we can transform the constraint \eqref{eq:P1-e} into a simpler form
\begin{equation}\label{eq:22}
	{\cal D}\left( {{{\mathbb P}_0}|{{\mathbb P}_1}} \right) \le 2{\epsilon ^2}.
\end{equation}
For simplicity of analysis, we define $z  \buildrel \Delta \over = \frac{{{\kappa _1}}}{{{\kappa _0}}}$ and then rewrite ${\cal D}\left( {{{\mathbb P}_0}|{{\mathbb P}_1}} \right)$ as a function of $z$ as $f\left( z  \right) = \ln \left( z  \right) + \frac{1}{z } - 1$, where $0< z \le 1$.
Since $\frac{{\partial f\left(z  \right)}}{{\partial z }} = \frac{{z  - 1}}{{{z ^2}}}$, $f\left(z  \right)$ is monotonically decreasing on the inverval $\left(0,1\right]$.
Based on this, the constraint \eqref{eq:22} can be further rewritten as
\begin{equation}\label{eq:23}
	0 < z  \le \Gamma,
\end{equation}
where $\Gamma$ can be found by solving the equation $\ln \left( \Gamma  \right) + \frac{1}{\Gamma}  - 1 = 2{\epsilon ^2}$ using the bisection or Newton's methods.
Since the inequality $z > 0$ is inherently held due to its definition, we can omit it for simplicity.

With the given receive filter $\bf w$ and the
constraint \eqref{eq:23}, the FDBF optimization problem can be expressed as
\begin{subequations}
    \begin{align}
        &\mathop {\max }\limits_{\{ {{\bf{v}}_{{\rm{FD}},i}}\} } \frac{{{{\left| {{\bf{h}}_{U + 2}^H{{\bf{v}}_{{\rm{FD}},U + 2}}} \right|}^2}}}{{\sum\limits_{i \in {\cal P}} {{{\left| {{\bf{h}}_{U + 2}^H{{\bf{v}}_{{\rm{FD}},i}}} \right|}^2}}  + \sigma _{\rm B}^2}},\label{eq:24-a}\\
        &\;\;\;{\rm{s}}.{\rm{t}}.\; \sum\limits_{i \in {\cal A}} {\left\| {{{\bf{v}}_{{\rm{FD}},i}}} \right\|_F^2} \le P,\\
        &\qquad\;\;\frac{{{{\left| {{\bf{h}}_u^H{{\bf{v}}_{{\rm{FD}},u}}} \right|}^2}}}{{\sum\limits_{i \in {\cal A},i \ne u} {{{\left| {{\bf{h}}_u^H{{\bf{v}}_{{\rm{FD}},i}}} \right|}^2}}  + \sigma _u^2}} \ge {2^{{\xi _u}}} - 1,\;u \in {\cal P},\\
        &\qquad\;\;\frac{{{{\left| {{\bf{h}}_{U + 1}^H{{\bf{v}}_{{\rm{FD}},U + 2}}} \right|}^2}}}{{\sum\limits_{i \in {\cal P}} {{{\left| {{\bf{h}}_{U + 1}^H{{\bf{v}}_{{\rm{FD}},i}}} \right|}^2}}  + \sigma _{U + 1}^2}} \le \Gamma  - 1,\\
        &\qquad\;\;\frac{{\sum\limits_{i \in {\cal A}} {{{\left| {{\varsigma _0}{{\bf{w}}^H}{\bf{A}}\left( {{\theta _0}} \right){{\bf{v}}_{{\rm{FD}},i}}} \right|}^2}} }}{{\sum\limits_{q \in {\cal Q}} {\sum\limits_{i \in {\cal A}} {{{\left| {{\varsigma _q}{{\bf{w}}^H}{\bf{A}}\left( {{\theta _q}} \right){{\bf{v}}_{{\rm{FD}},i}}} \right|}^2}} }  + \sigma _{\rm{r}}^2\left\| {\bf{w}} \right\|_F^2}} \ge \gamma,
    \end{align}\label{eq:24}%
\end{subequations}
where we define ${\bf{H}} = [ {{\bf{h}}_1}, \ldots ,{{\bf{h}}_U},{{\bf{h}}_{U + 1}},{{\bf{h}}_{U + 2}} ] = [ {{\bf{h}}_{{\rm{C}},1}},$ $ \ldots ,{{\bf{h}}_{{\rm{C}},U}},{{\bf{h}}_{\rm{W}}},{{\bf{h}}_{\rm{B}}} ] \in {{\mathbb C}^{{M_t} \times \left( {U + 2} \right)}}$, $\xi_{U+1} = \xi_{\rm W}$, $\sigma_u^2 = \sigma_{{\rm C},u}^2$ if $u\le U$, and $\sigma_u^2 = \sigma_{{\rm W}}^2$ if $u = U+1$  for simplicity of notation.

\subsubsection{Application of SDR Approach}\label{sec:III-C-2}
To solve the NP-hard problem \eqref{eq:24}, we apply the SDR approach to tackle it.
By defining ${{\bf{F}}_i} = {{\bf{v}}_i}{\bf{v}}_i^H, i \in {\cal A}$, we rewrite the problem \eqref{eq:24} as
\begin{subequations}
    \begin{align}
        &\mathop {\max }\limits_{\{ {{\bf{F}}_i}\} }\frac{{{\rm{Tr}}\left\{ {{{\bf{S}}_{U + 2}}{{\bf{F}}_{U + 2}}} \right\}}}{{\sum\limits_{i \in {\cal P}} {{\rm{Tr}}\left\{ {{{\bf{S}}_{U + 2}}{{\bf{F}}_i}} \right\}}  + \sigma _{\rm{B}}^2}},\\
        &\;\;{\rm{s.t.}}\;\sum\limits_{i \in {\cal A}} {{\rm{Tr}}\left\{ {{{\bf{F}}_i}} \right\}} \le P,\\
        &\qquad ( {{2^{{\xi _u}}} - 1} )( {\sum\limits_{i \in {\cal A}} {{\rm{Tr}}\left\{ {{{\bf{S}}_u}{{\bf{F}}_i}} \right\}}  + \sigma _{u}^2} ) \nonumber\\
        &\qquad\;\;- {2^{{\xi _u}}}{\rm{Tr}}\left\{ {{{\bf{S}}_u}{{\bf{F}}_u}} \right\} \le 0,\;u \in {\cal P},\\
        &\qquad {\rm{Tr}}\left\{ {{{\bf{S}}_{U + 1}}{{\bf{F}}_{U + 2}}} \right\} \nonumber\\
        &\qquad\;\; + ( {1 - \Gamma } )( {\sum\limits_{i \in {\cal P}} {{\rm{Tr}}\left\{ {{{\bf{S}}_{U + 1}}{{\bf{F}}_i}} \right\}}  + \sigma _{U + 1}^2} ) \le 0,\\
        &\qquad \gamma ( {\sum\limits_{q \in {\cal Q}} {\sum\limits_{i \in {\cal A}} {{\rm{Tr}}\left\{ {{{\bf{\Phi}}_q}{{\bf{F}}_i}} \right\}} }  + \sigma _{\rm{r}}^2\left\| {\bf{w}} \right\|_F^2} ) \nonumber\\
        &\qquad\;\; - \sum\limits_{i \in {\cal A}} {{\rm{Tr}}\left\{ {{{\bf{\Phi}}_0}{{\bf{F}}_i}} \right\}}  \le 0,\\
        &\qquad {{\bf{F}}_i}\succeq{\bf{0}},\;{\rm{rank}}\left\{ {{{\bf{F}}_i}} \right\} = 1,\;\forall i \in {\cal A},\label{eq:25-f}
    \end{align}\label{eq:25}%
\end{subequations}
where ${{\mathbf{\Phi }}_q} = {\left| {{\varsigma _q}} \right|^2}{{\mathbf{A}}^H}\left( {{\theta _q}} \right){\mathbf{w}}{{\mathbf{w}}^H}{\mathbf{A}}\left( {{\theta _q}} \right), q \in {\mathcal Q}$, ${{\bf{S}}_i} = {{\bf{h}}_i}{\bf{h}}_i^H,i \in {\cal A}$, and the constraint \eqref{eq:25-f} is equivalent to ${{\bf{F}}_i} = {{\bf{v}}_i}{\bf{v}}_i^H, i \in {\cal A}$.

To tackle the difficulty arising from the fractional programming (FP) problem \eqref{eq:25}, we leverage the Charnes-Cooper transformation \cite{cooper1962programming}, which introduces two variables
\begin{align}
\alpha  &= \frac{1}{{\sum\limits_{i \in {\cal P}} {{\rm{Tr}}\left\{ {{{\bf{S}}_{U + 2}}{{\bf{F}}_i}} \right\}}  + \sigma _{\rm{B}}^2}},\label{eq:26}\\
{{{\bf{\bar F}}}_i} &= \alpha {{\bf{F}}_i},\;\forall i \in {\cal A},\label{eq:27}
\end{align} 
and then relax the problem \eqref{eq:24} by neglecting the constraint ${\rm{rank}}\left\{ {{{\bf{F}}_i}} \right\} = 1,\;\forall i \in {\cal A}$, transforming the problem \eqref{eq:25} into
\begin{subequations}
    \begin{align}
        &\mathop {\max }\limits_{\{ {{{\bf{\bar F}}}_i}\} ,\alpha } \;{\rm{Tr}}\left\{ {{{\bf{S}}_{U + 2}}{{{\bf{\bar F}}}_{U + 2}}} \right\},\label{eq:28-a}\\
        &\;{\rm{s}}.{\rm{t}}.\;\;\;\sum\limits_{i \in {\cal A}} {{\rm{Tr}}\left\{ {{{{\bf{\bar F}}}_i}} \right\}}  \le \alpha P,\label{eq:28-b}\\
        &\qquad \left( {{2^{{\xi _u}}} - 1} \right)( {\sum\limits_{i \in {\cal A}} {{\rm{Tr}}\left\{ {{{\bf{S}}_u}{{{\bf{\bar F}}}_i}} \right\}}  + \alpha \sigma _{u}^2} )\nonumber \\
        &\qquad\;\;- {2^{{\xi _u}}}{\rm{Tr}}\left\{ {{{\bf{S}}_u}{{{\bf{\bar F}}}_u}} \right\} \le 0,\;u \in {\cal P},\label{eq:28-c}\\
        &\qquad {\rm{Tr}}\left\{ {{{\bf{S}}_{U + 1}}{{{\bf{\bar F}}}_{U + 2}}} \right\} \nonumber\\
        &\qquad\;\; + \left( {1 - \Gamma } \right)( {\sum\limits_{i \in {\cal P}} {{\rm{Tr}}\left\{ {{{\bf{S}}_{U + 1}}{{{\bf{\bar F}}}_i}} \right\}}  + \alpha \sigma _{U + 1}^2} ) \le 0, \label{eq:28-d}\\
        &\qquad \gamma ( {\sum\limits_{q \in {\cal Q}} {\sum\limits_{i \in {\cal A}} {{\rm{Tr}}\left\{ {{{\bf{\Phi}}_q}{{{\bf{\bar F}}}_i}} \right\}} }  + \alpha \sigma _{\rm{r}}^2\left\| {\bf{w}} \right\|_F^2} ) \nonumber\\
        &\qquad\;\;- \sum\limits_{i \in {\cal A}} {{\rm{Tr}}\left\{ {{{\bf{\Phi}}_0}{{{\bf{\bar F}}}_i}} \right\}}  \le 0,\label{eq:28-e}\\
        &\qquad \sum\limits_{i \in {\cal P}} {{\rm{Tr}}\left\{ {{{\bf{S}}_{U + 2}}{{{\bf{\bar F}}}_i}} \right\}}  + \alpha \sigma _{\rm{B}}^2 = 1,\label{eq:28-f}\\
        &\qquad \alpha  \ge 0, \;{{\bf{ \bar F}}_i}\succeq{\bf{0}},\;\forall i \in {\mathcal A},\label{eq:28-g}
    \end{align}\label{eq:28}%
\end{subequations}
which is a semi-definite programming (SDP) problem that can be easily solved using conventional methods, i.e., the interior-point method. 
After obtaining the optimal solution to \eqref{eq:26}, $\{ {{{\{ {{\bf{\bar F}}_i^ \star } \}}},{\alpha ^ \star }} \}$, we give the following proposition to derive the optimal solution to the FP problem \eqref{eq:25}.
\begin{prop}\label{prop:2}
    The FP problem \eqref{eq:25} is equivalent to the SDP problem \eqref{eq:28} via Charnes-Cooper transformation, and the optimal solution to \eqref{eq:28}, $\{{{\{ {{\bf{\bar F}}_i^ \star } \}_{i\in \cal A}}},{\alpha ^ \star } \}$, leads to the optimal solution to \eqref{eq:25}, ${\{ {{\bf{F}}_i^ \star } = {{{{\bf{\bar F}}_i^ \star } \mathord{\left/
 {\vphantom {{{\bf{\bar F}}_i^ \star } {{\alpha ^ \star }}}} \right.
 \kern-\nulldelimiterspace} {{\alpha ^ \star }}}} \}_{i\in \cal A}}$ that is of rank-1.
    \end{prop}
\begin{IEEEproof}
	Please refer to Appendix \ref{app:C}.
\end{IEEEproof}

Based on \textbf{Proposition \ref{prop:2}}, we can obtain the optimal SDR solution to the problem \eqref{eq:25}, $\{ {{\bf{F}}_i^ \star } \}$ that ${{\bf{F}}_i^ \star },i \in {\cal A}$ is of rank-1.
Accordingly, we can derive the optimal solution $\{ {{\bf{v}}_{{\rm{FD}},i}^ \star } \}$ to the FDBF design \eqref{eq:24}.

\subsection{Summary}

\setlength{\textfloatsep}{0.5em}
\begin{algorithm}[!t]
	\caption{The proposed FDBF design algorithm for the considered secure IoT system}
	\label{alg:1}
	\begin{algorithmic}[1]
		\STATE \textbf{Input:} System parameters.
  		\STATE \textbf{Output:} ${\bf w}$, ${\bf V}_{\rm FC}$, ${\bf v}_{\rm FW}$, and ${\bf v}_{\rm FB}$.
            \STATE \textbf{Initialization:} ${\bf V}_{\rm FC}^{[0]}$, ${\bf v}_{\rm FW}^{[0]}$, and ${\bf v}_{\rm FB}^{[0]}$, and $l = 0$.
		\WHILE{No Convergence}
		\STATE Update ${\bf w}^{[l]}$ by solving \eqref{eq:19}.
		\STATE Update $\{{\bf F}_i^{[l]}\}$ by solving \eqref{eq:23}.
            \STATE Update $\{ {{\bf{v}}_{{\rm{FD}},i}^{[l]}} \}$ by eigenvalue decomposition.
            \STATE Update $l = l + 1$.
		\ENDWHILE
        \STATE Extract ${\bf V}_{\rm FC}$, ${\bf v}_{\rm FW}$, and ${\bf v}_{\rm FB}$ from ${{\bf{V}}_{{\rm{FD}}}^{[l]}}$.
        \STATE ${\bf w} = {\bf w}^{[l]}$.
	\end{algorithmic}
\end{algorithm}

Based on the derivations above, the proposed FDBF design algorithm for the considered secure IoT system is summarized in \textbf{Algorithm \ref{alg:1}}.

Then, we analyze the complexity of \textbf{Algorithm \ref{alg:1}}.
Updating $\bf w$ needs the complexity of ${\cal O}\left( {M_t^3} \right)$.
The SDP problem \eqref{eq:26} is commonly solved by the interior-point method, so updating ${\{ {\bf F}_i\}}$ needs the complexity of ${\cal O}\left( {M_t^3\log \left( {1/\varepsilon}\right)} \right)$, where $\varepsilon$ is the precision of the interior-point method.
Besides, updating $\{ {{\bf{v}}_{{\rm{FD}},i}} \}$ by eigenvalue decomposition needs the complexity of ${\cal O}\left( { {U}M_t^3}\right)$.
To sum up, the overall complexity of \textbf{Algorithm \ref{alg:1}} is ${\cal O}\left( {M_t^3\log \left( {1/\varepsilon} \right) + {U}M_t^3} \right)$.

\section{HBF Deisign Algorithm}\label{sec:IV}

In this section, we consider the case where the HBF structure shown in Fig. \ref{fig:HBF} is implemented in the transmitter at Alice and propose a low-complexity algorithm to jointly design the transmit HBF and sensing receive filter.

\subsection{Problem Concretization under HBF Structure}

\subsubsection{Employment of HBF Structure}
We propose a transmitter in the HBF structure in Fig. \ref{fig:HBF}.
Specifically, the data streams ${{{\bf{s}}_{\rm{C}}},{s_{\rm{W}}},{s_{\rm{B}}}}$ are processed by the Carol digital beamformer ${{\bf{V}}_{\rm{DC}}} = \left[ {{{\bf{v}}_{{\rm{DC}},1}}, \ldots ,{{\bf{v}}_{{\rm{DC}},U}}} \right] \in {{\mathbb C}^{{M_t} \times U}}$, Willie digital beamformer ${{\bf{v}}_{\rm{DW}}} \in {{\mathbb C}^{{M_t}}}$, and Bob digial beamformer ${{\bf{v}}_{\rm{DB}}} \in {{\mathbb C}^{{M_t}}}$, respectively.
After DACs, the signals are up-converted to the RF domain via $N_t$ RF chains.
Then, the analog signals are modulated by the analog beamformer ${\bf F}_{\rm RF} \in {\mathbb C}^{M_t \times N_t}$ which is implemented by phase shifters.
Finally, the signals are emitted by $M_t$ antennas.
In this case, we substitute ${{\bf{V}}_{\rm{C}}} = [{{\bf{v}}_{{\rm{C}},1}},\ldots ,{{\bf{v}}_{{\rm{C}},U}} ] = [ {{\bf{V}}_{{\rm{RF}}}}{{\bf{v}}_{{\rm{DC}},1}}, \ldots ,{{\bf{V}}_{{\rm{RF}}}}{{\bf{v}}_{{\rm{DC}},U}} ]$, ${{\bf{v}}_{\rm{W}}} = {{\bf{V}}_{{\rm{RF}}}}{{\bf{v}}_{{\rm{DW}}}}$, ${{\bf{v}}_{\rm{B}}} = {{\bf{V}}_{{\rm{RF}}}}{{\bf{v}}_{{\rm{DB}}}}$ into the general form of the transmit signal \eqref{eq:1}.

\subsubsection{Problem Concretization}

Adopting the HBF structure at the transmitter, the unified design problem \eqref{eq:P0} can be concretized to a joint design of the analog beamformer, ${\bf V}_{\rm RF}$, digital beamformers, $\{ {{\bf{V}}_{{\rm{DC}}}} ,{{\bf{v}}_{{\rm{DW}}}},{{\bf{v}}_{{\rm{DB}}}}\}$, and the sensing receive filter, $\bf w$, which is given by
\begin{subequations}
    \begin{align}
        &\mathop {\max }\limits_{\scriptstyle\;\;{{\bf{V}}_{{\rm{RF}}}},{{\bf{V}}_{{\rm{DC}}}},\hfill\atop
        \scriptstyle{{\bf{v}}_{{\rm{DW}}}},{{\bf{v}}_{{\rm{DB}}}},{\bf{w}}\hfill} {\rm{Rate}}_{\rm{B}}\left({\bf V}_{\rm RF},{\bf V}_{\rm DC},{\bf v}_{\rm DW},{\bf v}_{\rm DB}\right),\label{eq:P2-a}\\
        &\quad\;\;{\rm{s.t.}}\left\| {{{\bf{V}}_{{\rm{RF}}}}{{\bf{V}}_{{\rm{DC}}}}} \right\|_F^2 + \left\| {{{\bf{V}}_{{\rm{RF}}}}{{\bf{v}}_{{\rm{DW}}}}} \right\|_F^2 + \left\| {{{\bf{V}}_{{\rm{RF}}}}{{\bf{v}}_{{\rm{DB}}}}} \right\|_F^2 \le P,\label{eq:P2-b}\\
        &\qquad \quad{\rm{Rat}}{{\rm{e}}_{{\rm{C}},u}}\left({\bf V}_{\rm RF},{\bf V}_{\rm DC},{\bf v}_{\rm DW},{\bf v}_{\rm DB}\right) \ge {\xi _u},u \in {\cal R},\label{eq:P2-d}\\
        &\qquad \quad{\rm{Rat}}{{\rm{e}}_{\rm{W}}}\left({\bf V}_{\rm RF},{\bf V}_{\rm DC},{\bf v}_{\rm DW},{\bf v}_{\rm DB}\right) \ge {\xi _{\rm{W}}},\label{eq:P2-e}\\
        &\qquad \quad{P_{\rm{E}}}\left({\bf V}_{\rm RF},{\bf V}_{\rm DC},{\bf v}_{\rm DW},{\bf v}_{\rm DB}\right) \le \epsilon,\label{eq:P2-f}\\
        &\qquad \quad{\rm{SIN}}{{\rm{R}}_{\rm{r}}}\left({\bf V}_{\rm RF},{\bf V}_{\rm DC},{\bf v}_{\rm DW},{\bf v}_{\rm DB},{\bf w}\right) \ge \gamma,\label{eq:P2-g}\\
        &\qquad \quad\left| {{{\bf{V}}_{{\rm{RF}}}}\left[ {m,n} \right]} \right| = 1,\forall m,n,\label{eq:P2-c}
    \end{align}\label{eq:P2}%
\end{subequations}
where the hardware constraint \eqref{eq:P2-c} denotes the constant modulus constraint of the analog beamformer.

Note that the problem \eqref{eq:P2} is non-convex due to the log-fractional expression in the objective function \eqref{eq:P2-a} and the constant modulus constraint \eqref{eq:P2-c}. 
To tackle these difficulties, we solve this problem in the next subsections.

\subsection{Receive Filter Optimization}
Similar to Section \ref{Sec-III-B-1}, the receiver filter optimization can be expressed by the following unconstrained problem
\begin{equation}\label{eq:31}
    \mathop {\max }\limits_{\bf{w}} \frac{{{{\bf{w}}^H}{\bf{\Xi }}_{\rm HBF}{\bf{w}}}}{{{{\bf{w}}^H}{\bf{\Lambda }}_{\rm HBF}{\bf{w}}}}
\end{equation}
where ${{\bf{\Xi }}_{{\rm{HBF}}}} = {\left| {{\varsigma _0}} \right|^2}{\bf{A}}\left( {{\theta _0}} \right){{\bf{V}}_{{\rm{RF}}}}{{\bf{V}}_{\rm{D}}}{\bf{V}}_{{\rm{RF}}}^H{\bf{V}}_{\rm{D}}^H{{\bf{A}}^H}\left( {{\theta _0}} \right)$, ${{\bf{\Lambda }}_{{\rm{HBF}}}} = \sum\nolimits_{q \in {\cal Q}} {{{\left| {{\varsigma _q}} \right|}^2}{\bf{A}}\left( {{\theta _q}} \right){{\bf{V}}_{{\rm{RF}}}}{{\bf{V}}_{\rm{D}}}{\bf{V}}_{{\rm{RF}}}^H{\bf{V}}_{\rm{D}}^H{{\bf{A}}^H}\left( {{\theta _q}} \right)}  + \sigma _{\rm{r}}^2{{\bf{I}}_{{M_r}}}$, and we define ${{\bf{V}}_{{\rm{D}}}} = \left[ {{{\bf{V}}_{{\rm{DC}}}},{{\bf{v}}_{{\rm{DW}}}},{{\bf{v}}_{{\rm{DB}}}}} \right] = \left[ {{{\bf{v}}_{{\rm{D}},1}}, \ldots ,{{\bf{v}}_{{\rm{D}},U + 2}}} \right] \in {{\mathbb C}^{{M_t} \times \left( {U + 2} \right)}}$ for simplicity of notation.
The generalized Rayleigh quotient problem \eqref{eq:31} can be solved using the same method in Section \ref{Sec-III-B-1}.

\subsection{HBF Optimization}\label{sec:IV-C}

\subsubsection{Problem Reformulation}
To tackle the difficulty arising from the log-fractional expression involved in the objective function \eqref{eq:P2-a}, we exploit the Rate-Weighted minimum mean square error (WMMSE) relationship in the following theorem.
\begin{theorem}\label{th:1}
    By introducing weighting coefficient $\omega$ and receive decoding coefficient $p$, the covert rate ${\rm{Rate}_{\rm{B}}}\left( {{{\bf{V}}_{{\rm{RF}}}},{{\bf{V}}_{\rm{D}}}} \right)$ can be equivalently rewritten as
    \begin{equation}\label{eq:32}
        {\rm{Rat}}{{\rm{e}}_{\rm{B}}}\left( {{{\bf{V}}_{{\rm{RF}}}},{{\bf{V}}_{\rm{D}}}} \right) =  - \omega {\cal E}\left( {{{\bf{V}}_{{\rm{RF}}}},{{\bf{V}}_{\rm{D}}},p} \right) + \log \omega  + 1,
    \end{equation}
    where MSE is defined by ${\cal E}( {{{\bf{V}}_{{\rm{RF}}}},{{\bf{V}}_{\rm{D}}},p} ) = {\mathbb E}\{ {\left| {{s_{\rm{B}}} - p{y_{\rm{B}}}} \right|} \}  = {\left| p \right|^2}( {\| {{\bf{h}}_{U + 2}^H{{\bf{V}}_{{\rm{RF}}}}{{\bf{V}}_{\rm{D}}}} \|_F^2 + \sigma _{\rm{B}}^2} ) - 2\Re \{ {p{\bf{h}}_{U + 2}^H{{\bf{V}}_{{\rm{RF}}}}{{\bf{v}}_{{\rm{D}},U + 2}}} \} + 1$.
\end{theorem}
\begin{IEEEproof}
	Please refer to \cite{shi2011iteratively}.
\end{IEEEproof}

Based on \textbf{Theorem \ref{th:1}}, we can equivalently transform the problem \eqref{eq:P2} into 
\begin{subequations}
    \begin{align}
        &\mathop {\min }\limits_{{{\bf{V}}_{{\rm{RF}}}},{{\bf{V}}_{\rm{D}}},p,\omega } \;\omega {\cal E}\left( {{{\bf{V}}_{{\rm{RF}}}},{{\bf{V}}_{\rm{D}}},p} \right) - \log \omega,\label{eq:33-a}\\
        &\qquad {\rm{s}}.{\rm{t}}.\qquad \eqref{eq:P2-b}-\eqref{eq:P2-g}.
    \end{align}\label{eq:33}%
\end{subequations}
Note that $\omega$ and $p$ only exist in the objective function \eqref{eq:33-a}, leading to the following unconstrained optimization problems.
\begin{align}
    {p^ \star } &= \mathop {\min }\limits_p \;{\cal E}\left( {{{\bf{V}}_{{\rm{RF}}}},{{\bf{V}}_{\rm{D}}},p} \right),\label{eq:34}\\
    {\omega ^ \star } &= \mathop {\min }\limits_\omega  \;\omega {\cal E}\left( {{{\bf{V}}_{{\rm{RF}}}},{{\bf{V}}_{\rm{D}}},p} \right) - \log \omega, \label{eq:35}
\end{align}
which can be solved by leveraging the Karush-Kuhn-Tucker (KKT) condition.
Specifically, the optimal solutions to the problems \eqref{eq:34} and \eqref{eq:35} are given by
\begin{align}
    p^ \star &= \frac{{{\bf{v}}_{{\rm{D}},U + 2}^H{\bf{V}}_{{\rm{RF}}}^H{{\bf{h}}_{U + 2}}}}{{\left\| {{\bf{h}}_{U + 2}^H{{\bf{V}}_{{\rm{RF}}}}{{\bf{V}}_{\rm{D}}}} \right\|_F^2 + \sigma _{\rm{B}}^2}},\label{eq:36}\\
    {\omega ^ \star } &= \frac{1}{{{\cal E}\left( {{{\bf{V}}_{{\rm{RF}}}},{{\bf{V}}_{\rm{D}}},p} \right)}}.\label{eq:37}
\end{align}

Applying \textbf{Lemma \ref{lemma:1}} again and substituting \eqref{eq:36} and \eqref{eq:37} to \eqref{eq:33-a}, the problem \eqref{eq:33} is more tractable than the problem \eqref{eq:P2}.
However, it is still difficult to solve due to the high coupling relationship between ${\bf F}_{\rm RF}$ and ${\bf F}_{\rm D}$ in the objective function and constraints \eqref{eq:P2-b}-\eqref{eq:P2-g}.
To decouple them and solve the constraints \eqref{eq:P2-b}-\eqref{eq:P2-g} from others, serval auxiliary variables ${\bf{Y}} = \left[ {{{\bf{y}}_1}, \ldots ,{{\bf{y}}_{U + 2}}} \right]$, ${{\bf{T}}_u} = \left[ {{{\bf{t}}_{u,1}}, \ldots ,{{\bf{y}}_{u,U + 2}}} \right],u \in {\cal P}$, ${\bf{G}} = \left[ {{{\bf{g}}_1}, \ldots ,{{\bf{g}}_{U + 2}}} \right]$ and ${\bf{M}} = \left[ {{{\bf{m}}_1}, \ldots ,{{\bf{m}}_{U + 2}}} \right]$ are introduced to further reformulate the problem \eqref{eq:33} as
\begin{subequations}
    \begin{align}
        &\mathop {\min }\limits_{\scriptstyle{\bf{Y}},\{ {{\bf{T}}_u}\} ,{\bf{G}},\hfill\atop
        \scriptstyle{\bf{M}},{{\bf{V}}_{{\rm{RF}}}},{{\bf{V}}_{\rm{D}}}\hfill} {\left| p \right|^2}\left\| {{\bf{h}}_{U + 2}^H{\bf{Y}}} \right\|_F^2 - 2\Re \left\{ {p{\bf{h}}_{U + 2}^H{{\bf{y}}_{U + 2}}} \right\},\label{eq:P3-a}\\
        &\qquad{\rm{s.t.}}\;\; \left\| {\bf{Y}} \right\|_F^2 \le P,\label{eq:P3-b}\\
        &\qquad \qquad\left| {{{\bf{V}}_{{\rm{RF}}}}\left[ {m,n} \right]} \right| = 1,\forall m,n,\label{eq:P3-c}\\
        &\qquad \qquad( {{2^{{\xi _u}}} - 1} )( {\sum\limits_{i \in {\cal A}} {{{\left| {{\bf{h}}_u^H{{\bf{t}}_{u,i}}} \right|}^2}}  + \sigma _u^2} ) \nonumber\\
        &\qquad\qquad\;\;- {2^{{\xi _u}}}{\left| {{\bf{h}}_u^H{{\bf{t}}_{u,u}}} \right|^2} \le 0,\;u \in {\cal P},\label{eq:P3-d}\\
        &\qquad\qquad{\left| {{\bf{h}}_{U + 1}^H{{\bf{g}}_{U + 2}}} \right|^2} \nonumber\\
        &\qquad\qquad\;\;+ ( {1 - \Gamma } )( {\sum\limits_{i \in {\cal P}} {{{\left| {{\bf{h}}_{U + 1}^H{{\bf{g}}_i}} \right|}^2}}  + \sigma _{\rm{W}}^2} ) \le 0,\label{eq:P3-e}\\
        &\qquad\qquad\gamma ( {\sum\limits_{q \in {\cal Q}} {\sum\limits_{i \in {\cal A}} {{{\left| {{\varsigma _q}{{\bf{w}}^H}{\bf{A}}( {{\theta _q}} ){{\bf{m}}_i}} \right|}^2}} }  + \sigma _{\rm{r}}^2\left\| {\bf{w}} \right\|_F^2} ) \nonumber\\
        &\qquad\qquad\;\;- \sum\limits_{i \in {\cal A}} {{{\left| {{\varsigma _0}{{\bf{w}}^H}{\bf{A}}( {{\theta _0}} ){{\bf{m}}_i}} \right|}^2}}  \le 0,\label{eq:P3-f}\\
        &\qquad\qquad{\bf{Y}} = {{\bf{T}}_u} = {\bf{G}} = {\bf{M}} = {{\bf{V}}_{{\rm{RF}}}}{{\bf{V}}_{\rm{D}}},u \in {\cal P},\label{eq:P3-g}
    \end{align}\label{eq:P3}%
\end{subequations}
Then, to tackle the equality constraint \eqref{eq:P3-g}, we penalize it into the objective function, leading to the following augmented Lagrangian (AL) minimization problem
\begin{subequations}
    \begin{align}
        &\mathop {\min }\limits_{\scriptstyle{\bf{Y}},\{ {{\bf{T}}_u}\} ,{\bf{G}},\hfill\atop
        \scriptstyle{\bf{M}},{{\bf{V}}_{{\rm{RF}}}},{{\bf{V}}_{\rm{D}}}\hfill} {\cal L}\left( {{\bf{Y}},\{ {{\bf{T}}_u}\} ,{\bf{G}},{\bf{M}},{{\bf{V}}_{{\rm{RF}}}},{{\bf{V}}_{\rm{D}}},{\bf{D}},\{ {{\bf{\Phi }}_u}\} ,{\bf{Z}},{\bf{\Omega }}} \right),\\
        &\qquad {\rm{s}}.{\rm{t}}.\qquad \eqref{eq:P3-b}-\eqref{eq:P3-f},
    \end{align}\label{eq:39}%
\end{subequations}
where the AL function is 
\begin{equation}
    \begin{aligned}
        &{\cal L}\left( {{\bf{Y}},\{ {{\bf{T}}_u}\} ,{\bf{G}},{\bf{M}},{{\bf{V}}_{{\rm{RF}}}},{{\bf{V}}_{\rm{D}}},{\bf{D}},\{ {{\bf{\Phi }}_u}\} ,{\bf{Z}},{\bf{\Omega }}} \right)\\
        & = {\left| p \right|^2}\left\| {{\bf{h}}_{U + 2}^H{\bf{Y}}} \right\|_F^2 - 2\Re \left\{ {p{\bf{h}}_{U + 2}^H{{\bf{y}}_{U + 2}}} \right\} \\
        & \;\; + \frac{{{\rho _1}}}{2}\left\| {{\bf{Y}} - {{\bf{V}}_{{\rm{RF}}}}{{\bf{V}}_{\rm{D}}} + {\bf{D}}} \right\|_F^2 + \frac{{{\rho _2}}}{2}\sum\limits_{u \in \cal P} {\left\| {{{\bf{T}}_u} - {\bf{Y}} + {{\bf{\Phi }}_u}} \right\|_F^2} \\
        &\;\; + \frac{{{\rho _3}}}{2}\left\| {{\bf{G}} - {\bf{Y}} + {\bf{Z}}} \right\|_F^2 + \frac{{{\rho _4}}}{2}\left\| {{\bf{M}} - {\bf{Y}} + {\bf{\Omega }}} \right\|_F^2,
    \end{aligned}\nonumber
\end{equation}
with penalty parameters $\rho_1, \rho_2, \rho_3, \rho_4 > 0$ and dual variables ${\bf{D}},\{ {{\bf{\Phi }}_u}\} ,{\bf{Z}},{\bf{\Omega }}$ for the constraints \eqref{eq:P3-b}-\eqref{eq:P3-f}.

\subsubsection{Alternating Optimization of AL Minimization Problem}

Based on the reformulations above, $( {\bf{Y}},\{ {{\bf{T}}_u}\} ,{\bf{G}},{\bf{M}},{{\bf{V}}_{{\rm{RF}}}},$ ${{\bf{V}}_{\rm{D}}},{\bf{D}},\{ {{\bf{\Phi }}_u}\} ,{\bf{Z}},{\bf{\Omega }} )$ can be updated alternatively by taking the following steps.

\textit{STEP 1 - Update $\bf Y$:}
With other variables fixed, $\bf Y$ can updated by solving
\begin{subequations}
    \begin{align}
        &\mathop {\min }\limits_{\bf{Y}} \;{\cal L}\left( {{\bf{Y}},\{ {{\bf{T}}_u}\} ,{\bf{G}},{\bf{M}},{{\bf{V}}_{{\rm{RF}}}},{{\bf{V}}_{\rm{D}}},{\bf{D}},\{ {{\bf{\Phi }}_u}\} ,{\bf{Z}},{\bf{\Omega }}} \right),\\
        &\;\;{\rm{s.t.}}\;\left\| {\bf{Y}} \right\|_F^2 \le P,
    \end{align}\label{eq:40}%
\end{subequations}
which can be equivalently rewritten as
\begin{subequations}
    \begin{align}
        &\mathop {\min }\limits_{{\bf{\tilde y}}} \;{{{\bf{\tilde y}}}^H}{\bf{B\tilde y}} - 2\Re \left\{ {{\bf{b}}_0^H{\bf{\tilde y}}} \right\} + \frac{{{\rho _1}}}{2}\left\| {{\bf{\tilde y}} - {\bf{\tilde v}} + {\bf{\tilde d}}} \right\|_F^2 \nonumber\\
        &\qquad+ \frac{{{\rho _2}}}{2}\sum\limits_{u \in {\cal P}} {\left\| {{{{\bf{\tilde t}}}_u} - {\bf{\tilde y}} + {{{\bm{\tilde \varphi }}}_u}} \right\|_F^2}  + \frac{{{\rho _3}}}{2}\left\| {{\bf{\tilde g}} - {\bf{\tilde y}} + {\bf{\tilde z}}} \right\|_F^2 \nonumber\\
        &\qquad+ \frac{{{\rho _4}}}{2}\left\| {{\bf{\tilde m}} - {\bf{\tilde y}} + {\bm{\tilde \omega }}} \right\|_F^2,\\
        & \;\;{\rm{s.t.}}\;\;{{{\bf{\tilde y}}}^H}{\bf{\tilde y}} \le P,
    \end{align}\label{eq:41}%
\end{subequations}
where ${\bf{B}} = {\left| p \right|^2}{{\bf{I}}_{U + 2}} \otimes {{\bf{S}}_{U + 2}}$, ${{\bf{b}}_0} = {\left[ {{{\bf{0}}_{1 \times {M_t}(U + 1)}},p{\bf{h}}_{U + 2}^H} \right]^H}$, ${\bf{\tilde v}} = {\rm{vec}}\left( {{{\bf{V}}_{{\rm{RF}}}}{{\bf{V}}_{\rm{D}}}} \right)$, ${\bf{\tilde y}} = {\rm{vec}}\left( {\bf{Y}} \right)$, ${{{\bf{\tilde t}}}_u} = {\rm{vec}}\left( {{{\bf{T}}_u}} \right),u \in {\cal P}$, ${\bf{\tilde g}} = {\rm{vec}}\left( {\bf{G}} \right)$, ${\bf{\tilde m}} = {\rm{vec}}\left( {\bf{M}} \right)$, ${\bf{\tilde d}} = {\rm{vec}}\left( {\bf{D}} \right)$, ${{{\bm{\tilde \varphi }}}_u} = {\rm{vec}}\left( {{{\bf{\Phi }}_u}} \right),u \in {\cal P}$, ${\bf{\tilde z}} = {\rm{vec}}\left( {\bf{Z}} \right)$, and ${\bm{\tilde \omega }} = {\rm{vec}}\left( {\bf{\Omega }} \right)$.
Note that \eqref{eq:41} is a quadratically constrained quadratic programming with one constraint (QCQP-1) problem, whose closed-form can be efficiently derived by leveraging the KKT condition.

\textit{STEP 2 - Update $\{ {{\bf{T}}_u}\}$:}
With other variables fixed, $\{ {{\bf{T}}_u}\}$ can be updated by solving
\begin{subequations}
    \begin{align}
        &\mathop {\min }\limits_{{{\bf{T}}_u}}\; \left\| {{{\bf{T}}_u} - {\bf{Y}} + {{\bf{\Phi }}_u}} \right\|_F^2,\\
        &\;\;{\rm{s.t.}}\;( {{2^{{\xi _u}}} - 1} )( {\sum\limits_{i \in {\cal A}} {{{\left| {{\bf{h}}_u^H{{\bf{t}}_{u,i}}} \right|}^2}}  + \sigma _{u}^2} ) - {2^{{\xi _u}}}{\left| {{\bf{h}}_u^H{{\bf{t}}_{u,u}}} \right|^2} \le 0,
    \end{align}\label{eq:42}%
\end{subequations}
which can be equivalently rewritten as 
\begin{equation}\label{eq:43}
    \mathop {\min }\limits_{{{{\bf{\tilde t}}}_u}} \;\left\| {{{{\bf{\tilde t}}}_u} - {\bf{\tilde y}} + {{{\bm{\tilde \varphi }}}_u}} \right\|_F^2,
    \;{\rm{s.t.}}\;{\bf{\tilde t}}_u^H{{\bf{Q}}_u}{{{\bf{\tilde t}}}_u} \le \left( {1 - {2^{{\xi _u}}}} \right)\sigma _u^2,
\end{equation}
where ${{\bf{Q}}_u} = - {2^{{\xi _u}}}{\rm{blkdiag}}\left( {{{\bf{0}}_{{M_t}\left( {u - 1} \right)}},{{\bf{S}}_u},{{\bf{0}}_{{M_t}\left( {U + 2 - u} \right)}}} \right) + \left( {{2^{{\xi _u}}} - 1} \right){{\bf{I}}_{U + 2}} \otimes {{\bf{S}}_u} , u \in \cal P$. 
The problem \eqref{eq:43} is a QCQP-1 problem that can be solved by leveraging the KKT condition.

\textit{STEP 3 - Update $\bf G$:}
With other variables fixed, $\bf G$ can be updated by solving
\begin{subequations}
    \begin{align}
        &\mathop {\min }\limits_{\bf{G}} \;\left\| {{\bf{G}} - {\bf{Y}} + {\bf{Z}}} \right\|_F^2,\\
        &\;\;{\rm{s.t.}}\;{\left| {{\bf{h}}_{U + 1}^H{{\bf{g}}_{U + 2}}} \right|^2} + ( {1 - \Gamma } )( {\sum\limits_{i \in {\cal P}} {{{\left| {{\bf{h}}_{U + 1}^H{{\bf{g}}_i}} \right|}^2}}  + \sigma _{U+1}^2} ) \le 0,
    \end{align}\label{eq:44}%
\end{subequations}
which can be equivalently rewritten as
\begin{equation}\label{eq:45}
    \mathop {\min }\limits_{{\bf{\tilde g}}} \;\left\| {{\bf{\tilde g}} - {\bf{\tilde y}} + {\bf{\tilde z}}} \right\|_F^2,\;{\rm{s.t.}}\;{{{\bf{\tilde g}}}^H}{\bf{R\tilde g}} \le \left( {\Gamma  - 1} \right)\sigma _{U+1}^2,
\end{equation}
where we define ${\bf{R}} = \left( {1 - \Gamma } \right){\rm{blkdiag}}\left( {{{\bf{I}}_{U + 1}} \otimes {{\bf{S}}_{U + 1}},{{\bf{0}}_{{M_t}}}} \right) + {\rm{blkdiag}}\left( {{{\bf{0}}_{{M_t}\left( {U + 1} \right)}},{{\bf{S}}_{U + 1}}} \right)$.
The problem \eqref{eq:45} is a QCQP-1 problem and can be solved by taking the KKT condition.

\textit{STEP 4 - Update $\bf M$:}
With other variables fixes, $\bf M$ can be updated by solving
\begin{subequations}
    \begin{align}
        &\mathop {\min }\limits_{\bf{M}} \;\left\| {{\bf{M}} - {\bf{Y}} + {\bf{\Omega }}} \right\|_F^2,\\
        &\;\;{\rm{s}}.{\rm{t}}.\;\;\gamma ( {\sum\limits_{q \in {\cal Q}} {\sum\limits_{i \in {\cal A}} {{{\left| {{\varsigma _q}{{\bf{w}}^H}{\bf{A}}( {{\theta _q}} ){{\bf{m}}_i}} \right|}^2}} }  + \sigma _{\rm{r}}^2\left\| {\bf{w}} \right\|_F^2} ) \nonumber\\
        &\qquad\;\; - \sum\limits_{i \in {\cal A}} {{{\left| {{\varsigma _0}{{\bf{w}}^H}{\bf{A}}( {{\theta _0}} ){{\bf{m}}_i}} \right|}^2}}  \le 0,
    \end{align}\label{eq:46}%
\end{subequations}
which can be equivalently rewritten as
\begin{equation}\label{eq:47}
    \mathop {\min }\limits_{{\bf{\tilde m}}} \;\left\| {{\bf{\tilde m}} - {\bf{\tilde y}} + {\bm{\tilde \omega }}} \right\|_F^2,\;{\rm{s.t.}}\;{{{\bf{\tilde m}}}^H}{\bf{J\tilde m}} \le  - \gamma \sigma _{\rm{r}}^2\left\| {\bf{w}} \right\|_F^2,
\end{equation}
where we define ${\bf{J}} =  - {\left| {{\varsigma _0}} \right|^2}{{\bf{I}}_{U + 2}} \otimes \left( {{{\bf{A}}^H}\left( {{\theta _0}} \right){\bf{w}}{{\bf{w}}^H}{\bf{A}}\left( {{\theta _0}} \right)} \right) + \sum\nolimits_{q \in {\cal Q}} {{{\left| {{\varsigma _q}} \right|}^2}{{\bf{I}}_{U + 2}} \otimes \left( {{{\bf{A}}^H}\left( {{\theta _q}} \right){\bf{w}}{{\bf{w}}^H}{\bf{A}}\left( {{\theta _q}} \right)} \right)}$.
Note that the problem \eqref{eq:46} is also a QCQP-1 problem, which can be solved by leveraging the KKT condition.

\textit{STEP 5 - Update ${{\bf{V}}_{{\rm{RF}}}}$:}
With other variables fixed, ${\bf F}_{\rm RF}$ can be updated by solving
\begin{subequations}
    \begin{align}
        &\mathop {\min }\limits_{{{\bf{V}}_{{\rm{RF}}}}} \;\left\| {{\bf{Y}} - {{\bf{V}}_{{\rm{RF}}}}{{\bf{V}}_{\rm{D}}} + {\bf{D}}} \right\|_F^2,\label{eq:48-a}\\
        &\;\;{\rm{s.t.}}\;\left| {{{\bf{V}}_{{\rm{RF}}}}\left[ {m,n} \right]} \right| = 1,\forall m,n.\label{eq:48-b}
    \end{align}\label{eq:48}%
\end{subequations}
To tackle the difficulty arising from the constant modulus constraint \eqref{eq:48-b}, we give the following proposition.
\begin{prop}\label{prop:3}
    The problem \eqref{eq:48} is a constant modulus constrained quadratic programming (CM-QP) problem.
    To tackle this, we propose a cyclic coordinate descent (CCD) framework to derive its closed-form solution.
    Specifically, the $\left(m,n\right)$-element at the $\left(k+1\right)$-th inner iteration is given by
    \begin{equation}\label{eq:49}
        {\bf{V}}_{{\rm{RF}}}^{\left[ {k + 1} \right]}\left[ {m,n} \right] =  - \exp \left( {\jmath \angle {{\bf{\Psi }}^{\left[ k \right]}}\left[ {m,n} \right]} \right)
    \end{equation}
    where ${{\bf{\Psi }}^{\left[ k \right]}} = \left( {{\bf{V}}_{{\rm{RF}}}^{\left[ k \right]}{{\bf{V}}_{\rm{D}}} - {\bf{Y}} - {\bf{D}}} \right){\bf{V}}_{\rm{D}}^H$.
\end{prop}
\begin{IEEEproof}
	Please refer to Appendix \ref{app:D}.
\end{IEEEproof}
Based on \textbf{Proposition \ref{prop:3}}, the solution to \eqref{eq:48} can be obtained until convergence in the CCD framework.

\textit{STEP 6 - Update ${{\bf{V}}_{\rm{D}}}$:}
With other variables fixed, ${\bf V}_{\rm D}$ can be updated by solving
\begin{equation}\label{eq:50}
    \mathop {\min }\limits_{{{\bf{V}}_{\rm{D}}}} \;\left\| {{\bf{Y}} - {{\bf{V}}_{{\rm{RF}}}}{{\bf{V}}_{\rm{D}}} + {\bf{D}}} \right\|_F^2,
\end{equation}
whose closed-form solution can be calculated by ${{\bf{V}}_{\rm{D}}} = {\left( {{\bf{V}}_{{\rm{RF}}}^H{{\bf{V}}_{{\rm{RF}}}}} \right)^{ - 1}}{\bf{V}}_{{\rm{RF}}}^H\left( {{\bf{Y}} + {\bf{D}}} \right)$.

\textit{STEP 7 - Update Dual Variables:}
With other variables fixed, the dual variables can be updated by
\begin{subequations}
    \begin{align}
        {{\bf{D}}^{\left[ {t + 1} \right]}} &= {{\bf{D}}^{\left[ t \right]}} + {{\bf{Y}}^{\left[ {t + 1} \right]}} - {\bf{V}}_{{\rm{RF}}}^{\left[ {t + 1} \right]}{\bf{V}}_{\rm{D}}^{\left[ {t + 1} \right]},\\
        {\bf{\Phi }}_u^{\left[ {t + 1} \right]} &= {\bf{\Phi }}_u^{\left[ t \right]} + {\bf{T}}_u^{\left[ {t + 1} \right]} - {{\bf{Y}}^{\left[ {t + 1} \right]}},u\in {\mathcal P}\\
        {{\bf{Z}}^{\left[ {t + 1} \right]}} &= {{\bf{Z}}^{\left[ t \right]}} + {{\bf{G}}^{\left[ {t + 1} \right]}} - {{\bf{Y}}^{\left[ {t + 1} \right]}},\\
        {{\bf{\Omega }}^{\left[ {t + 1} \right]}} &= {{\bf{\Omega }}^{\left[ t \right]}} + {{\bf{M}}^{\left[ {t + 1} \right]}} - {{\bf{Y}}^{\left[ {t + 1} \right]}}.
    \end{align}\label{eq:51}%
\end{subequations}

\vspace{-2em}
\subsection{Summary}

\setlength{\textfloatsep}{0.5em}
\begin{algorithm}[!t]
	\caption{The proposed HBF design algorithm for the considered secure IoT system}
	\label{alg:2}
	\begin{algorithmic}[1]
		\STATE \textbf{Input:} System parameters.
  	\STATE \textbf{Output:} ${\bf w}$, ${\bf V}_{\rm RF}$, ${\bf V}_{\rm DC}$, ${\bf v}_{\rm DW}$, and ${\bf v}_{\rm DB}$.
        \STATE \textbf{Initialization:} ${\bf V}_{\rm RF}^{[0]}$, ${\bf V}_{\rm FC}^{[0]}$, ${\bf v}_{\rm FW}^{[0]}$, and ${\bf v}_{\rm FB}^{[0]}$, and $l = 0$.
		\WHILE{No Convergence}
		  \STATE Update ${\bf w}^{[l]}$ by solving \eqref{eq:31}.
            \STATE Set $t = 0$.
        \WHILE{No Convergence}
            \STATE Update $p^{[t]}$ and $\omega^{[t]}$ by \eqref{eq:36} and \eqref{eq:37}, respectively.
            \STATE Update ${\bf Y}^{[t]}$ by solving \eqref{eq:41}.
            \STATE Update $\{ {{\bf{T}}_u^{[t]}}\}$ by solving \eqref{eq:43}.
            \STATE Update ${{\bf{G}}^{[t]}}$ by solving \eqref{eq:45}.
            \STATE Update ${{\bf{M}}^{[t]}}$ by solving \eqref{eq:47}.
            \STATE Update ${\bf{V}}_{{\rm{RF}}}^{[t]}$ by \eqref{eq:49}.
            \STATE Update ${{\bf{V}}_{\rm{D}}^{[t]}}$ by solving \eqref{eq:50}.
            \STATE Update dual variables by \eqref{eq:51}.
            \STATE $t = t + 1$.
        \ENDWHILE
            \STATE ${\bf{V}}_{{\rm{RF}}}^{[l]} = {\bf{V}}_{{\rm{RF}}}^{[t]}$, ${{\bf{V}}_{\rm{D}}^{[l]}} = {{\bf{V}}_{\rm{D}}^{[t]}}$.
            \STATE $l = l + 1$.
		\ENDWHILE
        \STATE ${\bf V}_{\rm RF} = {\bf{V}}_{{\rm{RF}}}^{[l]}$, and extract ${\bf V}_{\rm DC}$, ${\bf v}_{\rm DW}$, and ${\bf v}_{\rm DB}$ from ${{\bf{V}}_{{\rm{D}}}^{[l]}}$.
        \STATE ${\bf w} = {\bf w}^{[l]}$.
	\end{algorithmic}
\end{algorithm}

Based on the derivations above, the proposed HBF design algorithm for the considered secure IoT system is summarized in \textbf{Algorithm \ref{alg:2}}.

Next, we have an analysis of the complexity of \textbf{Algorithm \ref{alg:2}}.
Updating $\bf w$ needs the complexity of ${\cal O}\left( {M_t^3} \right)$.
Updating $\left( {\bf{Y}},\{ {{\bf{T}}_u}\} ,{\bf{G}} \right)$ needs the complexity of  ${\cal O}( {M_t^2{{U}^3}} )$. 
Updating ${\bf{M}}$ needs the complexity of ${\cal O}( {M_t^2{{U}^2} + M_t^3Q} )$. 
Updating ${\bf V}_{\rm RF}$ needs the complexity of ${\cal O}\left( {{N_{{\rm{CCD}}}}{M_t}N_t^2} \right)$, where $N_{\rm CCD}$ is the iteration number of the CCD method, and updating ${\bf V}_{\rm D}$ needs the complexity of ${\cal O}\left( N_t^2{M_t} + N_t^3 \right)$.
Therefore, the overall complexity of \textbf{Algorithm \ref{alg:2}} is ${\cal O}\left( {{N_{\rm O}}\left( {M_t^3 + {N_{\rm I}}\left( {M_t^2{U^3} + M_t^3Q + {N_{{\rm{CCD}}}}{M_t}N_t^2 + N_t^3} \right)} \right)} \right)$, where $N_{\rm O}$ and $N_{\rm I}$ denote the iteration number of the outer and inner loops, respectively.

\section{Extension to Beamforming Design with Imperfect Willie's CSI}

In this section, we extend the proposed beamforming design methods to a more challenging scenario where Willie is non-cooperative with Alice. 
In this case, there are no pilot signals or feedback links, and thus the perfect CSI of Willie is unavailable \cite{wang2021covert, pascual2005robust}.


\subsection{Problem Formulation}

The imperfect Willie's CSI is be modeled as \cite{pascual2005robust}
\begin{equation}\label{eq:52}
    {{\mathbf{h}}_{\rm{W}}} = {{{\mathbf{\hat h}}}_{\rm{W}}} + \Delta {{\mathbf{h}}_{\rm{W}}},\Delta {{\mathbf{h}}_{\rm{W}}} \in {\Omega _{\rm{W}}},
\end{equation}
where ${{{\mathbf{\hat h}}}_{\rm{W}}}$ is the estimated CSI vector of the Alice-to-Willie link, and $\Delta {{\mathbf{h}}_{\rm{W}}}$ is the CSI uncertainty within a spherical region
\begin{equation}\label{eq:53}
    {\Omega _{\rm{W}}} = \{ {\Delta {{\mathbf{h}}_{\rm{W}}}|\Delta {\mathbf{h}}_{\rm{W}}^H\Delta {{{\mathbf{ h}}}_{\rm{W}}} - {\delta ^2} \leqslant 0 }\}
\end{equation}

In this case, Willie is non-cooperative with Alice, thus, it is unnecessary for Alice to send signals to Willie.
Based on this setting, the QoS constraint (18d) should be removed, leading to the following unified problem
\begin{subequations}
    \begin{align}
        &\mathop {\max }\limits_{{{\bf{V}}_{{\rm{C}}}},{{\bf{v}}_{{\rm{W}}}},{{\bf{v}}_{{\rm{B}}}},{\bf{w}}} {\rm{Rate}}_{\rm{B}}\left({\bf V}_{\rm C},{\bf v}_{\rm W},{\bf v}_{\rm B}\right),\label{eq:P4-a}\\
        &\qquad\;{\rm{s.t.}}\quad\; \eqref{eq:P0-b},\eqref{eq:P0-c},\eqref{eq:P0-e}-\eqref{eq:P0-g},\eqref{eq:52}.\label{eq:P4-b}
    \end{align}\label{eq:P4}%
\end{subequations}
In the following subsections, based on two proposed methods in Sections \ref{sec:III} and \ref{sec:IV}, two robust FDBF and HBF algorithms are derived.

\subsection{Robust FDBF Design}

Substituting the Willie's CSI uncertainty \eqref{eq:52} into the transmission covertness constraint \eqref{eq:28-d}, we can obtain
\begin{equation}\label{eq:55}
    \begin{aligned}
            &{\mathbf{\hat h}}_{\rm{W}}^H{\mathbf{L}}{{{\mathbf{\hat h}}}_{\rm{W}}} + 2\Delta {\mathbf{h}}_{\rm{W}}^H{\mathbf{L}}{{{\mathbf{\hat h}}}_{\rm{W}}} \\
            &\quad+ \Delta {\mathbf{h}}_{\rm{W}}^H{\mathbf{L}}\Delta {{{\mathbf{\hat h}}}_{\rm{W}}} + \alpha\left( {1 - \Gamma } \right)\sigma _{\rm{W}}^2 \leqslant 0,
    \end{aligned}
\end{equation}
where ${\mathbf{L}} = {{{\mathbf{\bar F}}}_{U + 2}} + \left( {1 - \Gamma } \right)\sum\nolimits_{i \in \mathcal{P}} {{{{\mathbf{\bar F}}}_i}}$. 
Although \eqref{eq:55} are convex with the respect to the variables $\{{{{\mathbf{\bar F}}}_i}\}$, it involves semi-infinite constraints due to \eqref{eq:53}.
To tackle this difficulty, we give the following lemma to transform the constraints \eqref{eq:55} and \eqref{eq:53} into linear matrix inequalities (LMIs).

\begin{lemma}\label{lemma:2}
    Suppose functions $f_i({\bf s}),i\in\{1,2\},{\bf s}\in{\mathbb C}^{N}$, are defined as
    \begin{equation}
        {f_i}\left( {\mathbf{s}} \right) = {{\mathbf{s}}^H}{{\mathbf{A}}_i}{\mathbf{s}} + 2\Re \left\{ {{\mathbf{b}}_i^H{\mathbf{s}}} \right\} + {c_i},
    \end{equation}
    where ${{\mathbf{A}}_i}\in {\mathbb C}^{N \times N}$ is hermitian, ${\bf b}_i \in {\mathbb C}^N$, and $c_i \in {\mathbb R}$.
    Based on S-procedure \cite{boyd2004convex}, ${f_1}\left( {\mathbf{s}} \right) \leqslant 0 \Rightarrow {f_2}\left( {\mathbf{s}} \right) \leqslant 0$ holds if and only if there exists a variable $\eta \ge 0$ such that
    \begin{equation}
        \eta \left[ {\begin{array}{*{20}{c}}
  {{{\mathbf{A}}_1}}&{{{\mathbf{b}}_1}} \\ 
  {{\mathbf{b}}_1^H}&{{c_1}} 
\end{array}} \right] - \left[ {\begin{array}{*{20}{c}}
  {{{\mathbf{A}}_2}}&{{{\mathbf{b}}_2}} \\ 
  {{\mathbf{b}}_2^H}&{{c_2}} 
\end{array}} \right]{ \succeq } {\mathbf{0}}.
    \end{equation}
\end{lemma}
\begin{IEEEproof}
	Please refer to \cite{boyd2004convex}.
\end{IEEEproof}

Based on \textbf{Lemma \ref{lemma:2}}, constraints \eqref{eq:55} and \eqref{eq:53} can be converted into 
\begin{equation}\label{eq:58}
\left[ {\begin{array}{*{20}{c}}
  { - {\mathbf{L}} + \eta {{\mathbf{I}}_{{M_t}}}}&{ - {\mathbf{L}}{{{\mathbf{\hat h}}}_{\text{W}}}} \\ 
  { - {\mathbf{\hat h}}_{\rm{W}}^H{\mathbf{L}}}&{ - {\mathbf{\hat h}}_{\rm{W}}^H{\mathbf{L}}{{{\mathbf{\hat h}}}_{\rm{W}}} - \alpha \left( {1 - \Gamma } \right)\sigma _{\rm{W}}^2 - \eta {\delta ^2}} 
\end{array}} \right]{ \succeq } {\mathbf{0}}
\end{equation}
Then, the FDBF design problem with imperfect Willie's CSI can be reformulated as 
\begin{subequations}
    \begin{align}
        &\mathop {\max }\limits_{\{ {{{\mathbf{\bar F}}}_i}\} ,\alpha ,\eta } \;{\rm{Tr}}\left\{ {{{\mathbf{S}}_{U + 2}}{{{\mathbf{\bar F}}}_{U + 2}}} \right\},\\
        &\quad{\rm s.t.} \quad \eqref{eq:28-b},\eqref{eq:28}-\eqref{eq:28-g},\eqref{eq:58},\\
        &\qquad\quad \left( {{2^{{\xi _u}}} - 1} \right)( {\sum\limits_{i \in {\cal A}} {{\rm{Tr}}\left\{ {{{\bf{S}}_u}{{{\bf{\bar F}}}_i}} \right\}}  + \alpha \sigma _{u}^2} )\nonumber \\
        &\qquad\quad\;\;- {2^{{\xi _u}}}{\rm{Tr}}\left\{ {{{\bf{S}}_u}{{{\bf{\bar F}}}_u}} \right\} \le 0,\;u \in {\cal R},\\
        &\qquad\quad\;\;\eta \ge 0.
    \end{align}\label{eq:P5}%
\end{subequations}
Similar to the subsection \ref{sec:III-C-2}, we can solve the problem \eqref{eq:P5} and then derive the optimal FDBF. For the sake of brevity, we omit the detailed derivations.

\vspace{-0.5em}
\subsection{Robust HBF Design}

For the robust HBF design, \textbf{Lemma \ref{lemma:2}} is no longer applicable since solving the HBF design problem using SDP is non-trivial and computationally expensive.
To this end, we construct a sample of i.i.d. Alice-to-Willie channels approximate $\Omega _{\rm{W}}$ based on the set $\{ {{\mathbf{h}}_{{\rm{W}},k}} = {{{\mathbf{\hat h}}}_{\rm{W}}} + \Delta {{\mathbf{h}}_{{\rm{W}},k}}|{{{\mathbf{\hat h}}}_{\rm{W}}},\forall k \in \mathcal{K} = \left\{ {1, \ldots ,K} \right\} \}$, where $\Delta {{\mathbf{h}}_{{\rm{W}},k}} \in {\Omega _{\rm{W}}} = \{ {\Delta {{\mathbf{h}}_{\rm{W}}}|\left\| {\Delta {{\mathbf{h}}_{\rm{W}}}} \right\|_F^2 \leqslant {\delta ^2}} \}$ and $K$ is the sample size large enough.
Then, the HBF design problem with imperfect Willie's CSI can be approximately formulated as
\begin{subequations}
    \begin{align}
        &\mathop {\min }\limits_{\scriptstyle{\bf{Y}},\{ {{\bf{T}}_u}\} ,\{{\bf{G}}_k\},\hfill\atop
        \scriptstyle{\bf{M}},{{\bf{V}}_{{\rm{RF}}}},{{\bf{V}}_{\rm{D}}}\hfill} {\left| p \right|^2}\left\| {{\bf{h}}_{U + 2}^H{\bf{Y}}} \right\|_F^2 - 2\Re \left\{ {p{\bf{h}}_{U + 2}^H{{\bf{y}}_{U + 2}}} \right\},\\
        & \;\qquad{\rm{s.t.}}\quad \eqref{eq:P3-b},\eqref{eq:P3-c},\eqref{eq:P3-f},\\
        &\qquad \qquad\;\;( {{2^{{\xi _u}}} - 1} )( {\sum\limits_{i \in {\cal A}} {{{\left| {{\bf{h}}_u^H{{\bf{t}}_{u,i}}} \right|}^2}}  + \sigma _u^2} ) \nonumber\\
        &\qquad\qquad\quad- {2^{{\xi _u}}}{\left| {{\bf{h}}_u^H{{\bf{t}}_{u,u}}} \right|^2} \le 0,u \in {\cal R},\\
        &\qquad\qquad\;\;\left( {1 - \Gamma } \right)( {\sum\limits_{i \in \mathcal{P}} {{{\left| {{\mathbf{h}}_{{\text{W}},k}^H{{\mathbf{g}}_{k,i}}} \right|}^2}}  + \sigma _{\text{W}}^2} ) \nonumber\\
        &\qquad\qquad\quad + {\left| {{\mathbf{h}}_{{\text{W}},k}^H{{\mathbf{g}}_{k,U + 2}}} \right|^2} \leqslant 0,k \in \mathcal{K},\\
        &\qquad\qquad\;\;{\mathbf{Y}} = {{\mathbf{T}}_u} = {{\mathbf{G}}_k} = {\mathbf{M}} = {{\mathbf{V}}_{{\text{RF}}}}{{\mathbf{V}}_{\text{D}}},u \in \mathcal{R},k \in \mathcal{K},
    \end{align}\label{eq:P6}%
\end{subequations}
where auxiliary variables ${{\mathbf{G}}_k} = \left[ {{{\mathbf{g}}_{k,1}}, \ldots ,{{\mathbf{g}}_{k,U + 2}}} \right],k \in \mathcal{K}$.
Similar to the subsection \ref{sec:IV-C}, we can solve the problem \eqref{eq:P6} based on the alternating optimization method.
For the sake of brevity, we omit the detailed derivations.

\section{Numerical Simulations}

In this section, numerical simulations are provided to evaluate the performance of the proposed beamforming design for the physical layer covert transmission aided mmWave ISAC IoT system.

\vspace{-1em}
\subsection{System Parameters}

Unless specified otherwise, we assume Alice is equipped with $M_t = M_r = 32$ transmit/receive antennas.
It is assumed that the number of Carols is $U = 4$, and the number of data streams is $N_s = U + 2$ for all the communication UEs.
In the case of adopting the HBF structure at the transmitter, the number of RF chains is set as $N_t = N_s$.
We assume the transmit power at Alice is $P = 1{\rm W}$, the communication noise power is $\{\sigma_{{\rm C},u}^2 \}_{u = 1}^U = \sigma_{{\rm W}}^2 = \sigma_{{\rm B}}^2 = -5{\rm dBW}$, and the radar noise power is $\sigma_{\rm r}^2 = -10{\rm dBW}$.
Besides, there is a IoT sensing target located at $\theta_0 = 10^\circ$ and $Q = 2$ clutters located at $\theta_1 = -30^\circ, \theta_2 = 60^\circ$, and their complex amplitudes are assumed ${{{| {{\varsigma _0}} |}^2}} = 5{\rm dBW}$ and ${{{| {{\varsigma _q}} |}^2}} = 20{\rm dBW}, \forall q$, respectively.
The overt QoS threshold is set as $\{\xi_u\}_{u=1}^U = \xi_{\rm W} = \xi = 1\rm{bps/Hz}$, the transmission covertness requirement is set as $\epsilon = 0.001$, and the sensing SINR threshold is set as $\gamma = 10{\rm{dB}}$.
The sampling number of the angular range $[-90^\circ, 90^\circ]$ is $S = 181$.

\vspace{-1em}
\subsection{Benchmarks}

To highlight the potential of the proposed beamforming design in synergizing covert transmission and mmWave ISAC for the secure IoT system, we include the following benchmarks for comparison purposes:

\subsubsection{FDBF Comm. Only}
Alice with FDBF transmit structure only works for downlink communication. 
Specifically, the FDBF is obtained by solving the problem \eqref{eq:P1} without the constraint \eqref{eq:P1-f} using the same optimization method.

\subsubsection{HBF Comm. Only}
Alice with HBF transmit structure only works for downlink communication.
Specifically, the HBF is obtained by solving the problem \eqref{eq:P2} without the constraint \eqref{eq:P2-f} using the same optimization method.

\subsubsection{FDBF ZF ISAC \cite{yoo2006optimality}}
Alice with an FDBF transmit structure performs the zero-forcing (ZF) scheme to derive the beamformers for overt UEs by
    \begin{equation}
    {{\bf{V}}_{{\rm{CW}}}} = \frac{{{{\bf{H}}_{{\rm{CW}}}}{{( {{\bf{H}}_{{\rm{CW}}}^H{{\bf{H}}_{{\rm{CW}}}}} )}^{ - 1}}}}{{{{\| {{{\bf{H}}_{{\rm{CW}}}}{{( {{\bf{H}}_{{\rm{CW}}}^H{{\bf{H}}_{{\rm{CW}}}}} )}^{ - 1}}} \|}_F}}}\sqrt {P( {1 - \delta } )},
    \end{equation}
    where ${{\bf{V}}_{{\rm{CW}}}} = \left[ {{{\bf{V}}_{{\rm{FC}}}},{{\bf{v}}_{{\rm{FW}}}}} \right]$, ${{\bf{H}}_{{\rm{CW}}}} = \left[ {{{\bf{h}}_{{\rm{C}},1}}, \ldots ,{{\bf{h}}_{{\rm{C}},U}},{{\bf{h}}_{\rm{W}}}} \right]$, and $\delta$ denotes power allocation coefficient.
    Besides, the Bob FD beamformer ${\bf v}_{\rm B}$ is obtained by the proposed method.
    Specifically, ${\bf v}_{\rm B}$ is optimized iteratively by solving the problem \eqref{eq:P1} and adjusting $\delta$ to meet the constraints.

\subsubsection{FDBF MRT ISAC \cite{chen2005performance}}
Alice with an FDBF transmit structure performs the maximum ratio transmission (MRT) scheme to derive the beamformers for overt UEs by
    \begin{equation}
    {\bf{V}}_{{\rm{CW}}} = \frac{{{{\bf{H}}_{{\rm{CW}}}}}}{{{{\| {{{\bf{H}}_{{\rm{CW}}}}} \|}_F}}}\sqrt {P( {1 - \delta })}.
    \end{equation}
    Besides, the Bob FDBF ${\bf v}_{\rm B}$ is obtained by the proposed method.
    Specifically, ${\bf v}_{\rm B}$ is optimized in the same way as the above benchmark.

\subsubsection{TS HBF ISAC \cite{yu2016alternating}}
Alice with HBF structure designed by the two-stage (TS) method.
Specifically, the FDBF, ${{\mathbf{V}}_{{\rm{FD}}}}$, is first designed by the proposed FDBF method, and then the HBF is designed to approach this FDBF by solving
    \begin{equation}
    \begin{aligned}
        &\mathop {\min }\limits_{{{\mathbf{V}}_{{\rm{RF}}}},{{\mathbf{V}}_{\rm{D}}}} \left\| {{{\mathbf{V}}_{{\rm{RF}}}}{{\mathbf{V}}_{\rm{D}}} - {{\mathbf{V}}_{{\rm{FD}}}}} \right\|_F^2,\\
        &\quad{\rm{s.t.}}\;\;\;\left\| {{{\mathbf{V}}_{{\rm{RF}}}}{{\mathbf{V}}_{\rm{D}}}} \right\|_F^2 \leqslant P,\\
        &\qquad\quad\; \left| {{{\mathbf{V}}_{{\rm{RF}}}}\left[ {m,n} \right]} \right| = 1,\forall m,n,
    \end{aligned}
    \end{equation}
    which can be solved by the methods proposed in \cite{yu2016alternating}.

\vspace{-1em}
\subsection{Simulation Results}

\begin{figure}[!t]
\centering  
\includegraphics[width=0.7\linewidth]{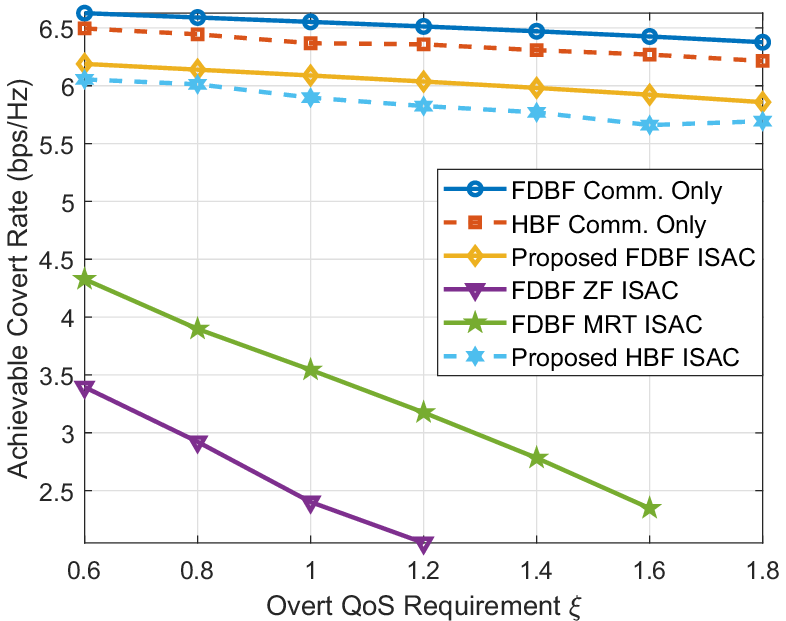}
\vspace{-1em}
\caption{Achievable covert rate versus overt QoS requirement $\xi$.}
\label{fig:1}
\vspace{-0.5em}
\end{figure}

\begin{figure}[!t]
\centering  
\includegraphics[width=0.7\linewidth]{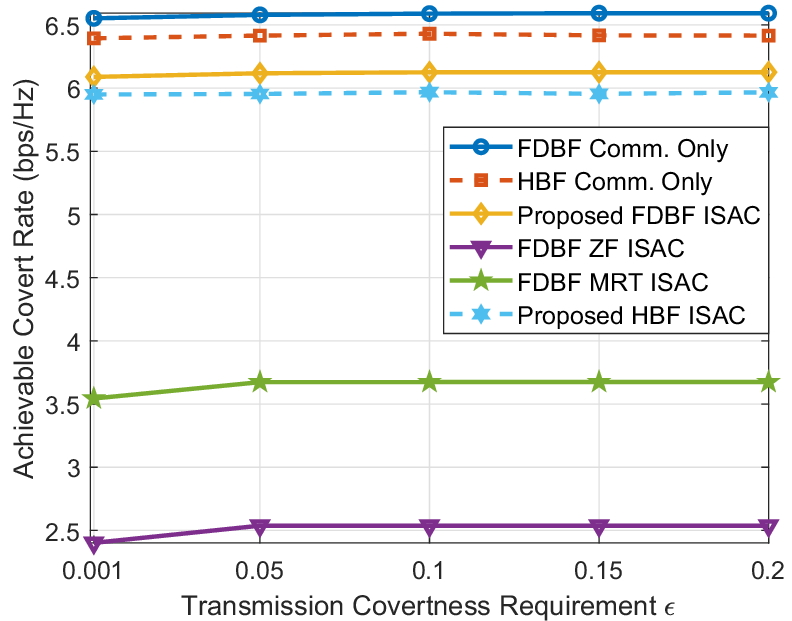}
\vspace{-1em}
\caption{Achievable covert rate versus covertness requirement $\epsilon$.}
\label{fig:2}
\end{figure}

Fig. \ref{fig:1} presents the achievable covert rate versus overt QoS requirement $\xi$. 
We can see that with $\xi$ increasing, the achievable covert rates obtained by the proposed FDBF and HBF ISAC design schemes decrease less violently than those obtained by the FDBF ZF and MRT ISAC schemes.
The covert rates achieved by the proposed FDBF and HBF ISAC schemes are significantly higher than those achieved by the FDBF ZF and MRT ISAC schemes.
Besides, the curves of the proposed ISAC schemes are close to those of the Comm. Only schemes in both FDBF and HBF structures, and the curve of the proposed HBF ISAC scheme approaches that of the proposed FDBF ISAC scheme.
This shows the effectiveness and superiority of the proposed design schemes in guaranteeing overt communication QoS and enhancing covert transmission.

Fig. \ref{fig:2} plots the achievable covert rate versus transmission covertness requirement $\epsilon$.
It can be observed that the covert rates achieved by the proposed design schemes are guaranteed almost the same with different $\epsilon$, while the covert rates achieved by the FDBF ZF and MRT ISAC schemes are noticeably lower when $\epsilon = 0.001$.
This demonstrates that the proposed design schemes can excellently guarantee the covert transmission in the IoT system at the cost of sacrificing very little covert communication QoS.

\begin{figure}[!t]
	\centering  
	\includegraphics[width=0.7\linewidth]{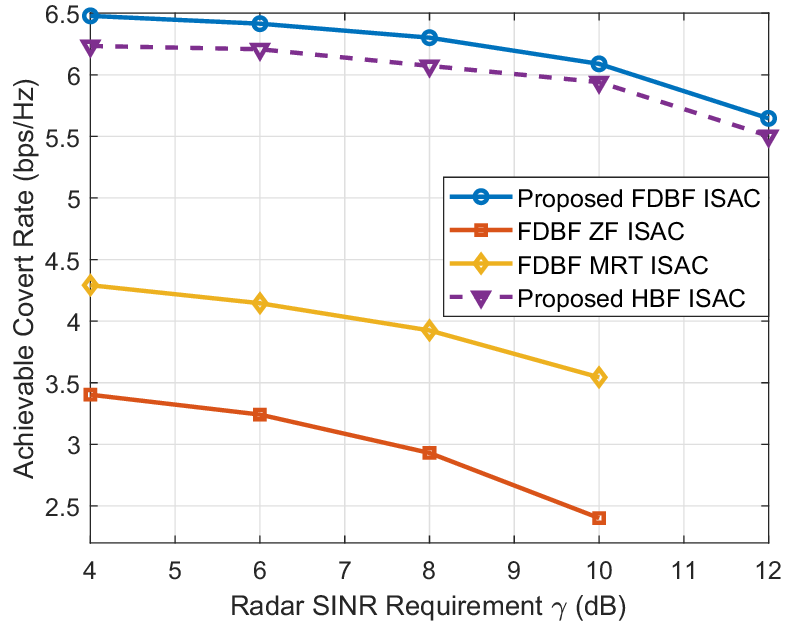}
 \vspace{-1em}
	\caption{Achievable covert rate versus sensing SINR requirement $\gamma$.}
	\label{fig:3-1}
 \vspace{-0.5em}
\end{figure}

Fig. \ref{fig:3-1} presents the achievable covert rate versus sensing SINR requirement $\gamma$.
It can be seen that the covert rates obtained by all the schemes decrease with the increase of $\gamma$, and the covert rates of the proposed design schemes are significantly higher than those of the FDBF ZF and MRT ISAC schemes.
Besides, we notice that the FDBF ZF and MRT ISAC schemes cannot work when $\gamma = 11,12{\rm dB}$.
This shows the effectiveness of the proposed mmWave ISAC in guaranteeing sensing performance and proves that a tradeoff exists between communication and sensing performance.

\begin{figure}[!t]
	\centering
	\subfigure[]{
		\includegraphics[width=0.485\linewidth]{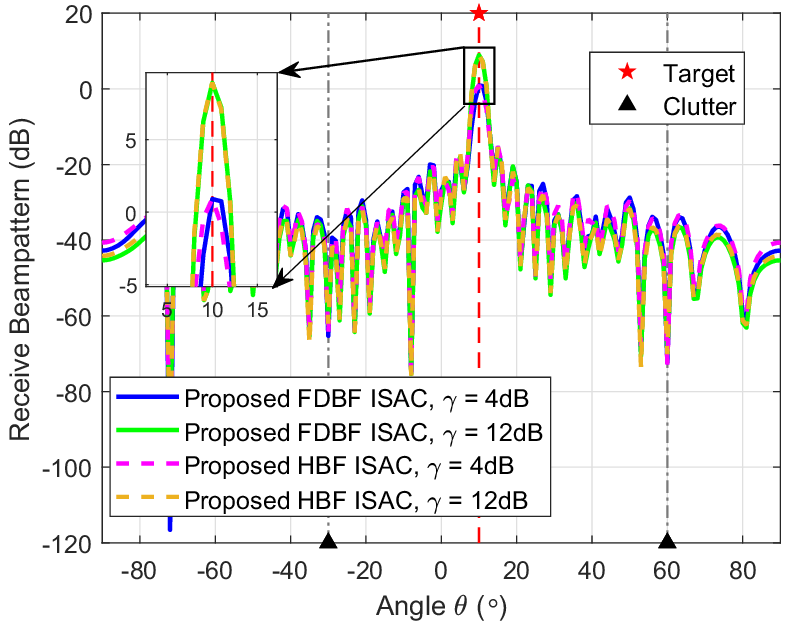} 
		\label{fig:3-2}	}
  \hspace{-1.35em}
	\subfigure[]{
		\includegraphics[width=0.485\linewidth]{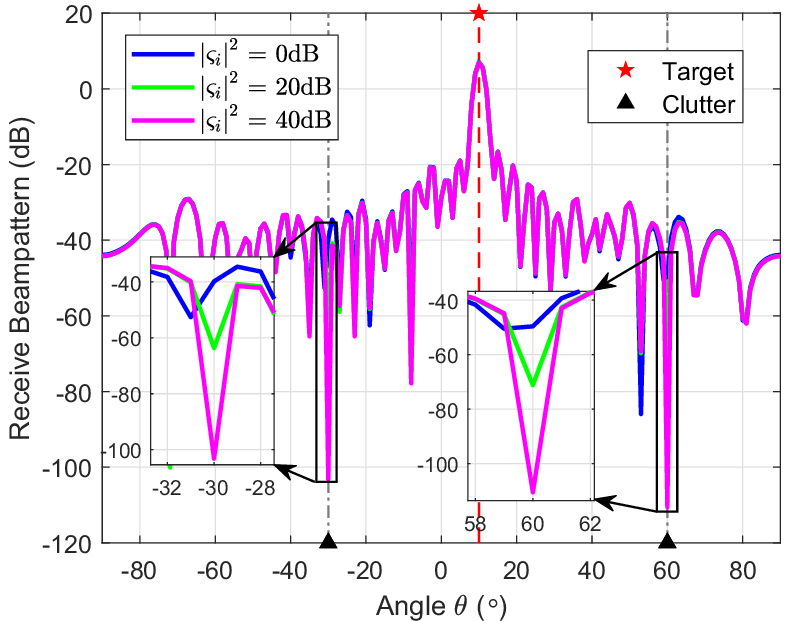} 
		\label{fig:3-3}	}
  \vspace{-1.5em}
	\caption{Receive Beampattern: (a) Different sensing SINR requirement $\gamma = 4,12{\rm dB}$; (b) Different clutter ${| {{\varsigma _i}} |^2} = 0,20,40{\rm dB}, \forall i$.}
\end{figure}



Fig. \ref{fig:3-2} presents the receive beampattern obtained by the proposed FDBF and HBF ISAC schemes with sensing SINR requirement $\gamma = 4,12{\rm dB}$.
We can observe that the transmit power is concentrated in the target direction and the notch is formed on the clutters.
The beampattern obtained by the FDBF ISAC scheme is almost the same as that by the HBF ISAC scheme\footnote{Considering that the receive beampatterns by the FDBF scheme and HBF are almost the same, we only plot one of them in the following for simplicity.} and realizes a higher mainlobe level with $\gamma = 12{\rm dB}$ than that with $\gamma = 4{\rm dB}$.
Besides, Fig. \ref{fig:3-3} presents the receive beampattern obtained by the proposed schemes with different clutters ${| {{\varsigma _q}} |^2} = 0,20,40{\rm dB}, \forall q$.
We see that the notch is deeper with higher clutter power.
This shows that the proposed schemes can achieve satisfactory beampattern behavior under different sensing requirements.

\begin{figure}[!t]
	\centering  
	\includegraphics[width=0.7\linewidth]{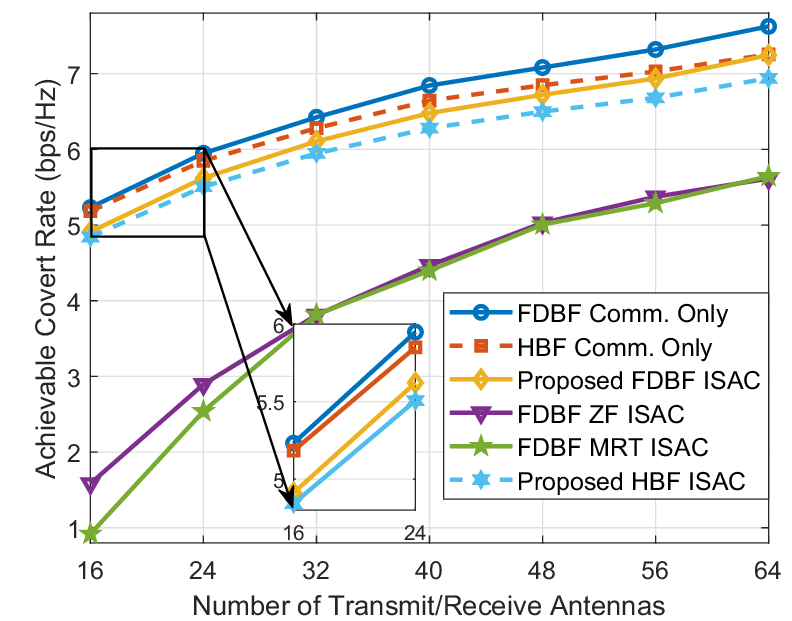}
 \vspace{-1em}
	\caption{Achievable covert rate versus number of transmit/receive antennas.}
	\label{fig:4-1}
     \vspace{-0.5em}
\end{figure}

Then, we investigate the impact of antennas and RF chains on the performance of the proposed mmWave ISAC IoT system. 
Fig. \ref{fig:4-1} plots the achievable covert rate versus the number of transmit/receive antennas.
As can be observed, with more antennas, the achievable covert rates obtained by all the schemes increase, and the covert rates achieved by the proposed schemes are always higher than those achieved by the FDBF ZF and MRT ISAC schemes.
This shows that higher numbers of antennas can improve the communication performance of the system.

\begin{figure}[!t]
	\centering  
	\includegraphics[width=0.7\linewidth]{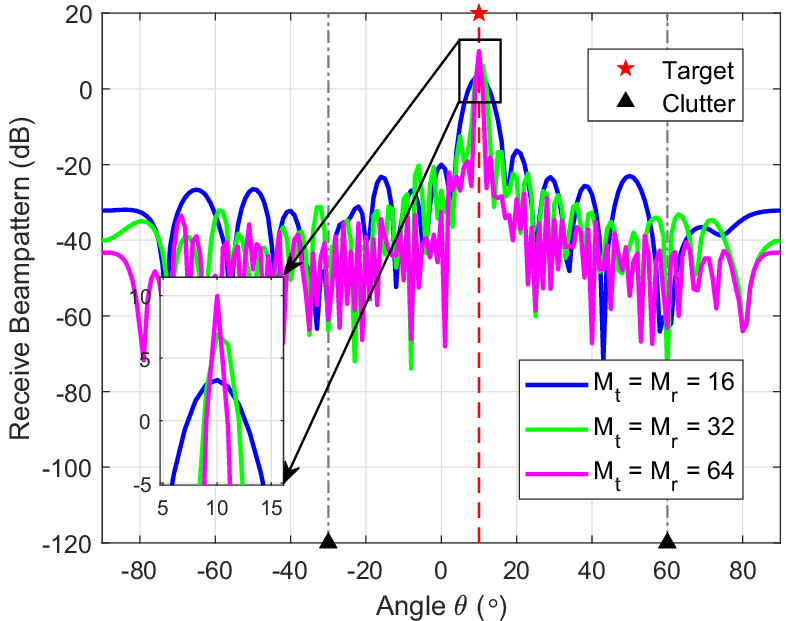}
 \vspace{-1em}
	\caption{Receive Beampattern with the number of transmit/receive antennas $M_t = M_r = 16,32,64$.}
	\label{fig:4-2}
\vspace{-0.5em}
\end{figure}

Fig. \ref{fig:4-2} presents the receive beampattern with different numbers of antennas $M_t = M_r = 16,32,64$.
We can see that the mainlobe level with 64 antennas is the highest, followed by the mainlobe level with 32 antennas in the second place, and the mainlobe level with 16 antennas is the lowest.
This demonstrates that higher numbers of antennas can enhance the radar performance of the systems to some extent.

\begin{figure}[!t]
	\centering  
	\includegraphics[width=0.7\linewidth]{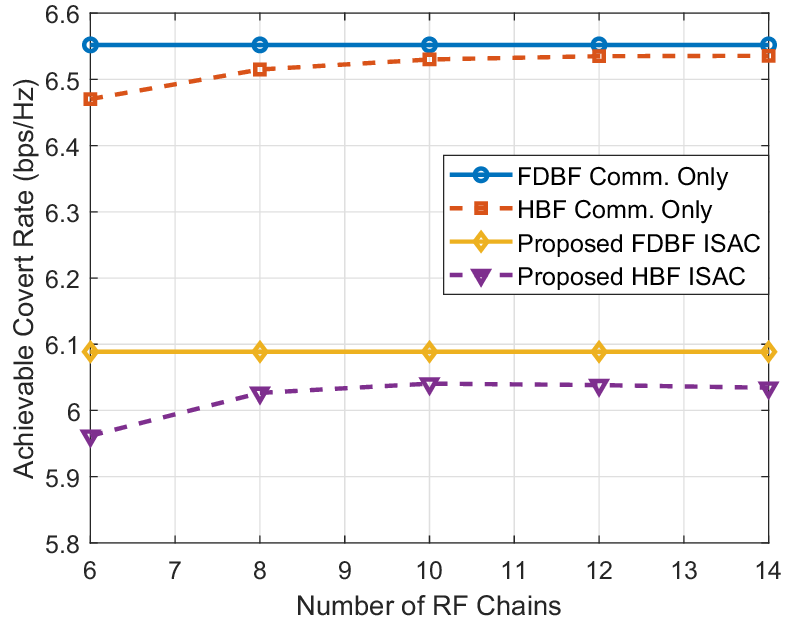}
 \vspace{-1em}
	\caption{Achievable covert rate versus number of RF chains.}
	\label{fig:4-3}
\end{figure}

Then, Fig. \ref{fig:4-3} plots the achievable covert rate versus the number of RF chains in the HBF structure.
It can be noticed that with the increasing number of RF chains, the covert rate achieved by the proposed HBF ISAC scheme rises and gets close to that achieved by the proposed FDBF ISAC scheme.
This validates the effectiveness of the proposed HBF ISAC design and gives an insight into the tradeoff between system performance and hardware costs.

\begin{figure}[!t]
	\centering  
	\includegraphics[width=0.7\linewidth]{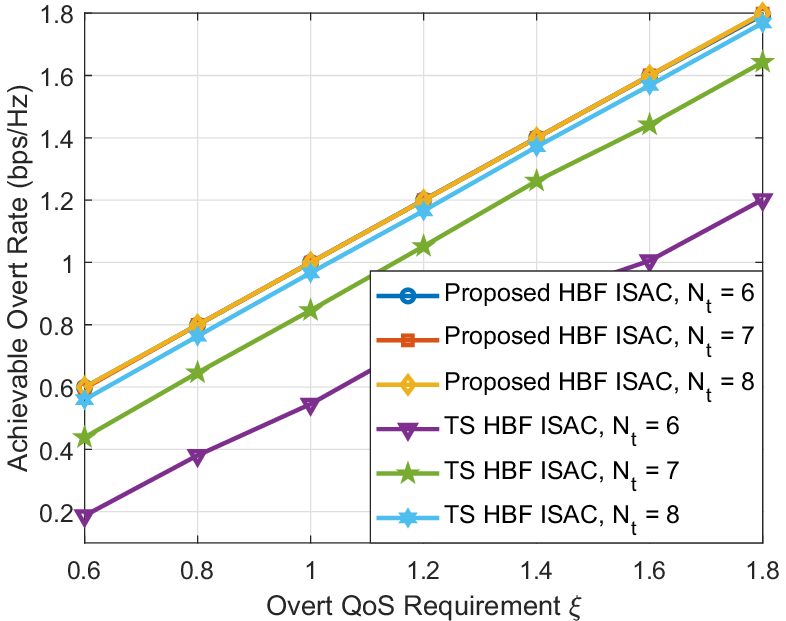}
 \vspace{-1em}
	\caption{Comparison of HBF algorithms: Achievable overt rate versus overt QoS requirement $\xi$.}
	\label{fig:5}
  \vspace{-0.5em}
\end{figure}

Moreover, we compare the proposed HBF ISAC algorithm with the TS HBF ISAC algorithm.
Fig. \ref{fig:5} presents the achievable overt rate versus overt QoS requirement $\xi$.
As expected, the overt rate achieved by the proposed HBF ISAC scheme equals to the required overt QoS.
However, the overt rate achieved by the TS HBF ISAC scheme is much lower than the required overt QoS when the number of RF chains is small.
This demonstrates the effectiveness and superiority of the proposed HBF algorithm in guaranteeing the overt QoS.

\begin{figure}[!t]
	\centering  
	\includegraphics[width=0.7\linewidth]{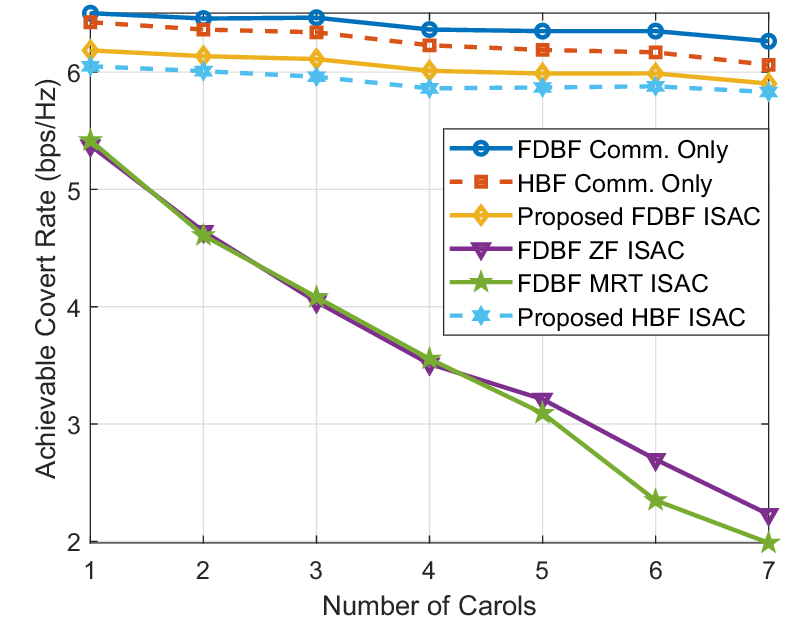}
 \vspace{-1em}
	\caption{Achievable covert rate versus number of Carols.}
	\label{fig:6}
\end{figure}

Fig. \ref{fig:6} plots the achievable covert rate versus the number of Carols.
It can be observed that as the number of Carols increases, the covert rates achieved by the proposed FDBF and HBF ISAC schemes are slightly decreasing while those achieved by the FDBF ZF and MRT ISAC schemes are dramatically decreasing.
This shows that the proposed covert transmission aided mmWave ISAC can guarantee satisfactory covert communication performance in multi-user scenarios.

Now, we explain the reasons behind the phenomena from simulation results above: 1) The benchmarks FDBF/HBF Comm. Only scheme only works for downlink communication, but there is a tradeoff between communication and sensing performance for ISAC systems \cite{liu2023seventy}; 2) The benchmarks FDBF ZF/MRT ISAC scheme are originated from a conventional precoding algorithm for MIMO communication systems, they are not the result of optimization making full use of the system resources; 3) The benchmark TS HBF ISAC scheme involves a two-stage optimization procedure, where once the FDBF matrix is fixed, the TS-based HBF matrices can only be optimized within the
restricted subspace, resulting in a reduction of the overall degree of freedom.

\begin{figure}[!t]
	\centering  
	\includegraphics[width=0.7\linewidth]{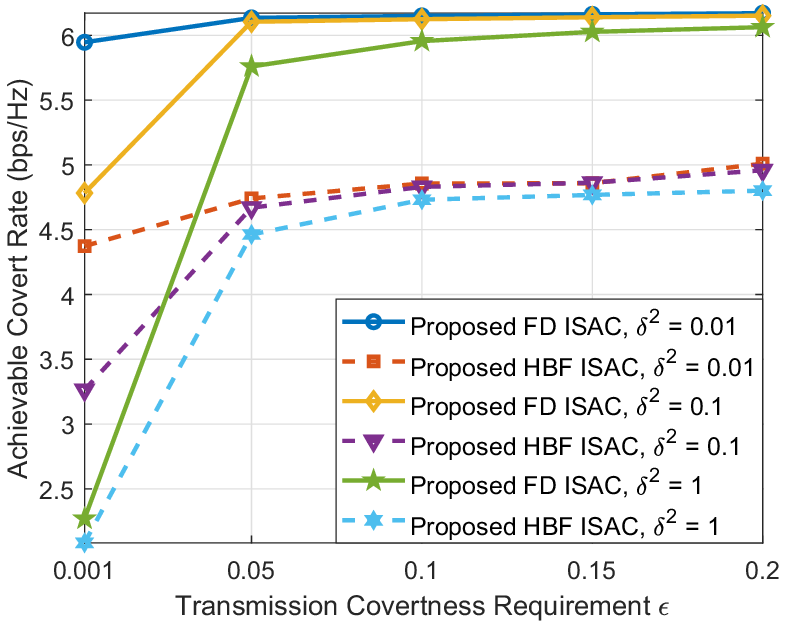}
 \vspace{-1em}
	\caption{Achievable covert rate versus covertness requirement $\epsilon$ with imperfect Willie's CSI.}
	\label{fig:7}
\end{figure}

Furthermore, we evaluate the performance of the proposed FDBF/HBF scheme with non-cooperative Willie.
We set the sample size for Alice-to-Willie channels is $K = 50$.
Fig. \ref{fig:7} plots the achievable covert rate versus covertness requirement $\epsilon$ with imperfect Willie's CSI, setting different channel uncertainty constraint $\delta^2 = 0.01,0.1,1$.
We can see that the covert rates achieved by the proposed FDBF and HBF ISAC schemes are lower with larger $\delta^2$.
Besides, the difference between the achievable covert rates is the largest with the smallest $\epsilon$.
This demonstrates the effectiveness of robust beamforming design and shows that the imperfect Willie's CSI has a larger impact on the covert rates at higher covertness requirements.

\section{Conclusion}


In this paper, we investigated how to synergize physical layer covert transmission and mmWave ISAC for secure IoT systems when adopting different transmit beamforming structures.
Specifically, we constructed the model of the considered secure IoT system involving multiple overt communication UEs, a covert communication UE, and an IoT sensing target in the presence of clutters.
To guarantee information transmission security, overt and covert communication QoS, and target detection accuracy, we devised a joint design scheme of transmit beamformer and receive filter and formulated a unified design problem by maximizing the covert rate while meeting the requirements of overt communication rate, transmission covertness and sensing SINR.
To solve the unified design problem in the specific cases of adopting FDBF and HBF structures, we proposed two algorithms: 1) a theoretical optimality-guaranteed algorithm based on the SDR method for the FDBF scheme; 2) a low-complexity algorithm based on alternating optimization method for the HBF scheme.
Furthermore, we extended the design into the scenario with the imperfect warden's CSI and proposed the corresponding robust beamforming design.
The simulation results demonstrate that the proposed FDBF and HBF algorithms outperform the conventional approaches.
Moreover, the FDBF scheme realizes superior performance in IoT security, communication and target detection performance, and the HBF scheme behaves better in hardware efficiency.
In future work, we can extend the current research into the scenario with multiple covert users or a wide-area detection requirement.


\appendices

\section{Proof of Proposition \ref{prop:1}}\label{app:A}
Substituting the likelihood functions in \eqref{eq:8} into the likelihood ratio test \eqref{eq:10}, we can obtain
\begin{equation}\label{eq:A1}
{\left| {{y_{\rm{W}}}} \right|^2} \underset{\mathcal{D}_0}{\overset{\mathcal{D}_1}{\gtrless} } \tau ,
\end{equation}
where $\tau = \frac{{{\kappa _0}{\kappa _1}}}{{{\kappa _1} - {\kappa _0}}}\ln \left( {\frac{{{\kappa _1}}}{{{\kappa _0}}}} \right)$ is the optimal detection threshold of Willie based on maximum-likelihood (ML) criterion \cite{poor2013introduction}.
For simplicity of notation, we define $t = {\left| {{y_{\rm{W}}}} \right|^2}$, and equivalently convert the likelihood ratio test \eqref{eq:8} into
\begin{equation}\label{eq:A2}
    \left\{ \begin{array}{l}
    {{\mathbb P}_0}:p\left( {t|{{\cal H}_0}} \right) = \frac{1}{{{\kappa _1}}}\exp \left( { - \frac{t}{{{\kappa _1}}}} \right),\\
    {{\mathbb P}_1}:p\left( {t|{{\cal H}_1}} \right) = \frac{1}{{{\kappa _0}}}\exp \left( { - \frac{t}{{{\kappa _0}}}} \right),
    \end{array} \right.
\end{equation}

Based on \eqref{eq:A2}, the false alarm probability and missed detection probability can be respectively calculated by
\begin{equation}
	\begin{aligned}
	\Pr \left( {{{\cal D}_1}|{{\cal H}_0}} \right) &= \int_\tau ^{ + \infty } {\frac{1}{{{\kappa _1}}}\exp \left( { - \frac{t}{{{\kappa _1}}}} \right)dt} = \exp \left( { - \frac{\tau }{{{\kappa _0}}}} \right) \\
	& = {\left( {\frac{{{\kappa _1}}}{{{\kappa _0}}}} \right)^{ - \frac{{{\kappa _1}}}{{{\kappa _1} - {\kappa _0}}}}},\\
	\Pr \left( {{{\cal D}_0}|{{\cal H}_1}} \right)&= \int_0^\tau  {\frac{1}{{{\kappa _1}}}\exp \left( { - \frac{t}{{{\kappa _1}}}} \right)dt }  = 1 - \exp \left( { - \frac{\tau }{{{\kappa _1}}}} \right)\\
	& = 1 - {\left( {\frac{{{\kappa _1}}}{{{\kappa _0}}}} \right)^{ - \frac{{{\kappa _0}}}{{{\kappa _1} - {\kappa _0}}}}}.
	\end{aligned}\nonumber
\end{equation}

The proof is completed.

\section{Proof of Lemma \ref{lemma:2}}\label{app:B}

Based on NP test, an equality relationship between the total detection error probability and the total variation distance between ${\mathbb P}_0$ and ${\mathbb P}_1$, ${{\cal V}_T}\left( {{{\mathbb P}_0},{{\mathbb P}_1}} \right)$, can be given by \cite{lehmann1986testing}
\begin{equation}\label{eq:D1}
	{P_{\rm{E}}} = 1 - {{\cal V}_T}\left( {{{\mathbb P}_0},{{\mathbb P}_1}} \right),
\end{equation}
where ${{\cal V}_T}\left( {{{\mathbb P}_0},{{\mathbb P}_1}} \right) = \frac{1}{2}{\left\| {p\left( {{y_{\rm{W}}}|{{\cal H}_0}} \right) - p\left( {{y_{\rm{W}}}|{{\cal H}_1}} \right)} \right\|_F}$.
Based on Pinsker's inequality \cite{cover1999elements}, we also have
\begin{equation}\label{eq:D2}
	{{\cal V}_T}\left( {{{\mathbb P}_0},{{\mathbb P}_1}} \right) \le \sqrt {\frac{1}{2}{\cal D}\left( {{{\mathbb P}_0}|{{\mathbb P}_1}} \right)}, 
\end{equation}
where the relative entropy ${\cal D}\left( {{{\mathbb P}_0}|{{\mathbb P}_1}} \right)$ is defined as
\begin{equation}\label{eq:D3}
	{\cal D}\left( {{{\mathbb P}_0}|{{\mathbb P}_1}} \right) = \int_{ - \infty }^{ + \infty } {p\left( {{y_{\rm{W}}}|{{\cal H}_0}} \right)\ln \frac{{p\left( {{y_{\rm{W}}}|{{\cal H}_0}} \right)}}{{p\left( {{y_{\rm{W}}}|{{\cal H}_1}} \right)}}d{y_{\rm{W}}}}.
\end{equation}

Then, we substitute \eqref{eq:8} into \eqref{eq:D3}, leading to
\begin{equation}\label{eq:D4}
	\begin{aligned}
	    &D\left( {{{\mathbb P}_0}|{{\mathbb P}_1}} \right) \\
	    =& \int_{ - \infty }^{ + \infty }{\frac{1}{{\pi {\kappa _0}}}\exp \left( { - \frac{{{{\left| {{y_{\rm{W}}}} \right|}^2}}}{{{\kappa _0}}}} \right)\ln \left( {\frac{{\frac{1}{{ {\kappa _0}}}\exp \left( { - \frac{{{{\left| {{y_{\rm{W}}}} \right|}^2}}}{{{\kappa _0}}}} \right)}}{{\frac{1}{{ {\kappa _1}}}\exp \left( { - \frac{{{{\left| {{y_{\rm{W}}}} \right|}^2}}}{{{\kappa _1}}}} \right)}}} \right)d{y_{\rm{W}}}} \\
	    =& \ln \left( {\frac{{{\kappa _1}}}{{{\kappa _0}}}} \right) + \frac{{{\kappa _0}}}{{{\kappa _1}}} - 1.
	\end{aligned}
\end{equation}

Combing \eqref{eq:D1} and \eqref{eq:D2}, we can obtain the lower bound of $P_{\rm E}$, ${{\hat P}_{\rm{E}}}$ as follows.
\begin{equation}\label{eq:D5}
{P_{\rm{E}}} \ge 1 - \sqrt {\frac{1}{2}D\left( {{{\mathbb P}_0}|{{\mathbb P}_1}} \right)}  \buildrel \Delta \over = {{\hat P}_{\rm{E}}}.
\end{equation}

The proof is completed.

\section{Proof of Proposition \ref{prop:2}}\label{app:C}

Applying the Charnes-Cooper transformation to the problem \eqref{eq:25} and dropping the non-convex rank-1 constraint, we can relax the problem \eqref{eq:25} to the SDP problem \eqref{eq:28} with defining ${{{\bf{\bar F}}}_i} = \alpha {{\bf{F}}_i},\;\forall i \in {\cal A}$ and $\alpha > 0$ as follows.
\begin{subequations}
    \begin{align}
        &\mathop {\max }\limits_{\{ {{{\bf{\bar F}}}_i}\} ,\alpha } \;{\rm{Tr}}\left\{ {{{\bf{S}}_{U + 2}}{{{\bf{\bar F}}}_{U + 2}}} \right\},\label{eq:C1-a}\\
        &\;{\rm{s}}.{\rm{t}}.\;\;\;\sum\limits_{i \in {\cal A}} {{\rm{Tr}}\left\{ {{{{\bf{\bar F}}}_i}} \right\}}  \le \alpha P,\label{eq:C1-b}\\
        &\qquad \left( {{2^{{\xi _u}}} - 1} \right)( {\sum\limits_{i \in {\cal A}} {{\rm{Tr}}\left\{ {{{\bf{S}}_u}{{{\bf{\bar F}}}_i}} \right\}}  + \alpha \sigma _{u}^2} )\nonumber \\
        &\qquad\;\;- {2^{{\xi _u}}}{\rm{Tr}}\left\{ {{{\bf{S}}_u}{{{\bf{\bar F}}}_u}} \right\} \le 0,\;u \in {\cal P},\label{eq:C1-c}\\
        &\qquad {\rm{Tr}}\left\{ {{{\bf{S}}_{U + 1}}{{{\bf{\bar F}}}_{U + 2}}} \right\} \nonumber\\
        &\qquad\;\; + \left( {1 - \Gamma } \right)( {\sum\limits_{i \in {\cal P}} {{\rm{Tr}}\left\{ {{{\bf{S}}_{U + 1}}{{{\bf{\bar F}}}_i}} \right\}}  + \alpha \sigma _{U + 1}^2} ) \le 0,\label{eq:C1-d}\\
        &\qquad \gamma ( {\sum\limits_{q \in {\cal Q}} {\sum\limits_{i \in {\cal A}} {{\rm{Tr}}\left\{ {{{\bf{T}}_q}{{{\bf{\bar F}}}_i}} \right\}} }  + \alpha \sigma _{\rm{r}}^2\left\| {\bf{w}} \right\|_F^2} ) \nonumber\\
        &\qquad\;\;- \sum\limits_{i \in {\cal A}} {{\rm{Tr}}\left\{ {{{\bf{T}}_0}{{{\bf{\bar F}}}_i}} \right\}}  \le 0,\label{eq:C1-e}\\
        &\qquad \sum\limits_{i \in {\cal P}} {{\rm{Tr}}\left\{ {{{\bf{S}}_{U + 2}}{{{\bf{\bar F}}}_i}} \right\}}  + \alpha \sigma _{\rm{B}}^2 = 1,\label{eq:C1-f}\\
        &\qquad {{\bf{ \bar F}}_i}\succeq{\bf{0}},\;\forall i \in {\mathcal A},\label{eq:C1-g}
    \end{align}\label{eq:C1}%
\end{subequations}
which can be solved by the conventional methods, i.e., the interior-point method. 
Note that $\alpha  \ge 0$ is removed because it is simplicity guaranteed by the constraint \eqref{eq:C1-b}.
The optimal solution to \eqref{eq:25} can be obtained by ${\{ {{\bf{F}}_i^ \star } = {{{{\bf{\bar F}}_i^ \star } \mathord{\left/
 {\vphantom {{{\bf{\bar F}}_i^ \star } {{\alpha ^ \star }}}} \right.
 \kern-\nulldelimiterspace} {{\alpha ^ \star }}}} \}_{i\in \cal A}}$, where the optimal solution to \eqref{eq:28} is 
$\{{{\{ {{\bf{\bar F}}_i^ \star } \}_{i\in \cal A}}},{\alpha ^ \star } \}$.

Denoting the optimal objective value of \eqref{eq:C1} as $f^{\star}$, we have
\begin{subequations}
    \begin{align}
        &\mathop {\min }\limits_{\{ {{{\bf{\bar F}}}_i}\} ,\alpha } \;\;\sum\limits_{i \in {\cal A}} {{\rm{Tr}}\left\{ {{{{\bf{\bar F}}}_i}} \right\}}  - \alpha P,\label{eq:C2-a}\\
        &\quad{\rm{s}}.{\rm{t}}.\;\;\;{\rm{Tr}}\left\{ {{{\bf{S}}_{U + 2}}{{{\bf{\bar F}}}_{U + 2}}} \right\} \ge {f^ \star },\label{eq:C2-b}\\
        &\qquad\quad \eqref{eq:C1-c}-\eqref{eq:C1-g},
    \end{align}\label{eq:C2}%
\end{subequations}
whose optimal solution is the same as the problem \eqref{eq:C1}.
Otherwise, there would be another solution ${\{ {\bf{\bar F}}_i^\prime \} _{i \in A}}$ that can achieve ${\rm{Tr}}\{ {{{\bf{S}}_{U + 2}}{\bf{\bar F}}_{U + 2}^\prime } \} > {f^ \star}$.
However, it is impossible because the objective value \eqref{eq:C2-a} can be further decreased until the equality ${\rm{Tr}}\{ {{{\bf{S}}_{U + 2}}{\bf{\bar F}}_{U + 2}^\prime } \} = {f^ \star}$ is reached.
Then, based on \cite{huang2009rank}, the optimal solution to \eqref{eq:C2} follows
\begin{equation}\label{eq:C3}
    \sum\limits_{i \in {\cal A}} {{{\left( {{\rm{rank}}\left\{ {{\bf{\bar F}}_i^ \star } \right\}} \right )}^2}}  + {\left( {{\rm{rank}}\left\{ \alpha  \right\}} \right)^2} \le U + 5.
\end{equation}
Since ${\bf{\bar F}}_i^ \star  \ne {\bf{0}},\forall i \in {\cal A}$ and $\alpha^{\star} \ge 0$, we can derive ${\rm{rank}}\{ {{\bf{\bar F}}_i^ \star } \} = 1,\forall i \in {\cal A}$.
Therefore, the rank-1 optimal solution can be guaranteed though the rank-1 constraint is removed.

The proof is completed.

\section{Proof of Proposition \ref{prop:3}}\label{app:D}

Given other variables, ${\bf F}_{\rm RF}$ can be updated by
\begin{subequations}
    \begin{align}
        &\mathop {\min }\limits_{{{\bf{V}}_{{\rm{RF}}}}} \;\left\| {{\bf{Y}} - {{\bf{V}}_{{\rm{RF}}}}{{\bf{V}}_{\rm{D}}} + {\bf{D}}} \right\|_F^2,\label{eq:E-1-a}\\
        &\;\;{\rm{s}}.{\rm{t}}.\;\;\left| {{{\bf{V}}_{{\rm{RF}}}}\left[ {m,n} \right]} \right| = 1,\forall m,n,\label{eq:E-1-b}
    \end{align}\label{eq:E-1}%
\end{subequations}
To solve the problem above, we apply the Majorization-Minimization (MM) method, the upper-bound of the objective function \eqref{eq:E-1-a} at the $\left(k+1\right)$-th inner iteration is given by
\begin{equation}
    \begin{aligned}
        &\| {{\bf{Y}} - {\bf{V}}_{{\rm{RF}}}^{\left[ {k + 1} \right]}{{\bf{V}}_{\rm{D}}} + {\bf{D}}} \|_F^2 \\
        &\le \| {{\bf{Y}} - {\bf{V}}_{{\rm{RF}}}^{\left[ k \right]}{{\bf{V}}_{\rm{D}}} + {\bf{D}}} \|_F^2 \\
        &\quad - \Re \{ {{\rm{Tr}}\{ {{{( {( {{\bf{Y}} - {\bf{V}}_{{\rm{RF}}}^{\left[ k \right]}{{\bf{V}}_{\rm{D}}} + {\bf{D}}} ){\bf{V}}_{\rm{D}}^H} )}^H}( {{{\bf{V}}_{{\rm{RF}}}} - {\bf{V}}_{{\rm{RF}}}^{\left[ k \right]}} )} \}} \}\\
        &= \Re \{ {{\rm{Tr}}\{ {{\bf{V}}_{{\rm{RF}}}^H( {{\bf{V}}_{{\rm{RF}}}^{\left[ k \right]}{{\bf{V}}_{\rm{D}}} - {\bf{Y}} - {\bf{D}}} ){\bf{V}}_{\rm{D}}^H} \}} \} + C,
    \end{aligned}\nonumber
\end{equation}
where $C = - \Re \{ {{\rm{Tr}}\{ {{{( {{\bf{V}}_{{\rm{RF}}}^{\left[ k \right]}} )}^H}( {{\bf{V}}_{{\rm{RF}}}^{\left[ k \right]}{{\bf{V}}_{\rm{D}}} - {\bf{Y}} - {\bf{D}}} ){\bf{V}}_{\rm{D}}^H} \}} \} + \| {{\bf{Y}} - {\bf{V}}_{{\rm{RF}}}^{\left[ k \right]}{{\bf{V}}_{\rm{D}}} + {\bf{D}}} \|_F^2$.
Then, ${\bf{V}}_{{\rm{RF}}}^{\left[ {k + 1} \right]}$ can be updated by
\begin{equation}\label{eq:E-3}
    \begin{aligned}
        &\mathop {\min }\limits_{{\bf{V}}_{{\rm{RF}}}^{\left[ {k + 1} \right]}} \Re \{ {{\rm{Tr}}\{ {{{( {{\bf{V}}_{{\rm{RF}}}^{\left[ {k + 1} \right]}} )}^H}{{\bf{\Psi }}^{\left[ k \right]}}} \}} \},\\
        &\;\;\;{\rm{s.t.}}\;\left| {{{\bf{V}}_{{\rm{RF}}}}\left[ {m,n} \right]} \right| = 1,\forall m,n,
    \end{aligned}
\end{equation}
where ${{\bf{\Psi }}^{\left[ k \right]}} = ( {{\bf{V}}_{{\rm{RF}}}^{\left[ k \right]}{{\bf{V}}_{\rm{D}}} - {\bf{Y}} - {\bf{D}}} ){\bf{V}}_{\rm{D}}^H$.
The closed-form solution to \eqref{eq:E-3} is ${\bf{V}}_{{\rm{RF}}}^{\left[ {k + 1} \right]}\left[ {m,n} \right] =  - \exp \left( {\jmath \angle {{\bf{\Psi }}^{\left[ k \right]}}\left[ {m,n} \right]} \right)$.

The proof is completed.

\balance
\bibliographystyle{IEEEtran}
\bibliography{IEEEabrv,ref_abrv.bib}

\clearpage

\end{document}